\documentclass[twocolumn,
                showpacs,
                nofootinbib,
                nobibnotes,
                aps,
                superscriptaddress,
                amssymb,
                amsmath,
                floatfix,
                reprint,
                prx,
                notitlepage,
                showkeys,
                10pt,
		longbibliography,
		]{revtex4-2}

\usepackage{amssymb}
\usepackage{amsmath}
\usepackage{cprotect}
\usepackage{graphicx}
\usepackage[caption=false]{subfig}
\usepackage{dcolumn}
\usepackage{url}
\usepackage{dsfont}
\usepackage{xcolor}
\usepackage[ruled]{algorithm2e}
\usepackage{bbold}
\usepackage[binary-units=true]{siunitx}
\usepackage{glossaries}
\usepackage{pgf}
\usepackage{pgfplots}
\usepgfplotslibrary{groupplots}
\pgfplotsset{compat=1.18}
\usepackage{tikz}
\usepackage{bm}
\usepackage{hhline}
\usepackage{spverbatim}
\usepackage{mhchem}
\usepackage{multirow}
\usepackage{afterpage}
\usepackage[english]{babel}

\usepackage{float}
\usepackage{placeins}
\usepackage{flafter}

\usepackage[colorlinks=true,breaklinks=true,allcolors=blue]{hyperref}
\usepackage[normalem]{ulem}

\makeatletter
\let\cat@comma@active\@empty
\makeatother

\newcommand{\Fig}[1]{Fig.~\ref{fig:#1}}
\newcommand{\Tab}[1]{Tab.~\ref{tab:#1}}
\newcommand{\App}[1]{App.~\ref{app:#1}}

\begin{document}
\title{Scalable Neural Wave Functions via Information Mixing and Hierarchical Network Architectures}

\author{Faith Oyedemi}
\email{foyedemi@uottawa.ca}
\affiliation{Department of Physics, University of Ottawa, Ontario, K1N 6N5, Canada}
\author{Stefanie Czischek}
\email{stefanie.czischek@uottawa.ca}
\affiliation{Department of Physics, University of Ottawa, Ontario, K1N 6N5, Canada}

\date{\today}

\begin{abstract}
Numerically simulating quantum many-body systems is one of the central challenges in modern physics, as the exponential growth of the Hilbert space with system size renders large-scale exact solutions intractable. Over the past decade, generative artificial neural networks have emerged as powerful wave function ans\"atze that have powered advances in finding ground state representations, reconstructing quantum states from data, and modeling dynamics in quantum many-body systems. While recurrent neural networks (RNNs) are computationally efficient, the recurrent nature can limit their capacity to capture long-range correlations. Transformer models with their underlying self-attention mechanism excel at capturing these long-range correlations, but incur increased computational requirements, restricting applications to small physical systems. In this work, we address these limitations by exploring the multi-layer perceptron (MLP) mixer model as a wave function ansatz. By mixing information globally and locally through the system via a computationally lightweight mixing strategy, this architecture promises to combine the ability to capture long-range correlations with low computational costs. We benchmark the model on finding ground state representations of two-dimensional quantum many-body systems in various phases of matter and investigate the behavior of representational accuracy and computational efficiency for different physical system sizes. By comparing its performance to RNNs and transformer models with single-site and patched input definitions, we demonstrate that the MLP mixer architecture performs competitively with state-of-the-art network models. We further show that a hierarchical combination of this model with the transformer model outperforms other state-of-the-art architectures for large-scale quantum systems, opening new avenues for investigating quantum phases of matter.
\end{abstract}
\maketitle

\section{Introduction}
Theoretical studies of systems of interacting quantum particles remain a daunting task owing to the exponential scaling of the Hilbert space with system size~\cite{RN82}.
This renders the exact treatment of quantum many-body systems intractable.
Yet, such systems are fundamental models underlying quantum computation~\cite{RN59, RN60, RN72}, quantum information~\cite{RN65, RN66, RN68, RN73}, quantum chemistry~\cite{RN69, RN74, RN75}, or quantum condensed matter physics~\cite{RN76, RN77, RN79, RN82}, emphasizing the need for their detailed theoretical understanding.

The variational principle has motivated approximative numerical studies of quantum many-body systems using parameterized wave functions and variational optimizations of free parameters~\cite{RN83, RN84, RN85, RN86, RN87, RN88}.
Early efforts such as the Jastrow, density-density, backflow, or Gutzwiller wave function representations rely on physically motivated, handcrafted ans\"atze~\cite{RN39, RN40, RN82}.
Such trial wave functions capture specific correlation structures in the physical system, but lack expressivity, limiting the accuracy of their representations~\cite{RN85, DMRG, RN80, RN82}.

A more general wave function ansatz uses artificial neural networks (ANNs) as parameterized representations, where the network weights and biases act as variational parameters~\cite{RN80, SharirEtAl}.
This approach has proven its powers in finding ground state representations of given Hamiltonians, reconstructing quantum states from projective measurement data, and modeling real-time dynamics in closed and open quantum systems~\cite{Lange_2024}.

The strength and computational efficiency of ANN-based wave function representations depend strongly on the underlying ANN architecture.
While early works use restricted Boltzmann machines as wave function ansatz~\cite{Jia_2019}, a new state of the art has been achieved by using recurrent neural networks (RNNs) instead~\cite{Hibat-a, Hibat-b}.
Inspired by natural language processing, RNNs in an autoregressive setup have been used to model quantum many-body systems as sequences of particles, providing normalized outputs that can be interpreted as wave function amplitudes~\cite{Hibat-a, SharirEtAl}.

In natural language processing, RNNs have nowadays mainly been replaced by transformer architectures that reveal novel powers in modeling sequential data with strong long-range correlations, resulting in the introduction of large language models~\cite{vaswani2017attention}.
Motivated by this new standard, transformer models have since found applications in quantum many-body physics where they overcome limitations of RNN-based representations when being used as a wave function ansatz~\cite{Lange_2024}.
However, the novel strength in capturing long-range correlations comes at the cost of long computational run times that scale quadratically with the quantum system size, eliminating studies of large-scale systems~\cite{Sprague&Czischek, RN142}.
While the computational efficiency has been significantly improved by considering sequences of patches of quantum particles rather than sequences of individual particles~\cite{trockman2022patches, Sprague&Czischek, Viteritti}, further reductions in computational run times are needed to efficiently model systems at scales that are required for investigating quantum phases of matter~\cite{Sprague&Czischek, RN142, RN143, Viteritti}.

In this work, we approach this problem by introducing a wave function ansatz based on the multi-layer perceptron (MLP) mixer model~\cite{tolstikhin2021mlpmixer}.
This ANN architecture is entirely constructed from feed-forward operations, making it computationally efficient while attaining competitive results with state-of-the-art transformer models in image processing applications~\cite{tolstikhin2021mlpmixer}.
Here, we tailor the MLP mixer architecture to be used as a wave function ansatz and variationally optimize it to find ground state representations of two-dimensional quantum many-body systems.
Considering the two-dimensional transverse-field Ising model (TFIM) and two-dimensional arrays of Rydberg atoms, we compare the performance of the MLP mixer model with RNN and transformer models, also investigating patched input arrangements for all three models.
We compare representational accuracy and computational efficiency over all models in different phases of matter and in the vicinity of quantum phase transitions and show that, while RNNs are the most accurate and efficient models for small-scale quantum systems, the MLP mixer architecture outperforms the RNN and the transformer model when scaling up to large system sizes.
With this, we introduce a wave function ansatz that promises novel insights into quantum condensed matter physics by enabling efficient and accurate studies of large-scale, higher-dimensional quantum many-body systems.

\section{Artificial Neural Network Architectures}

\begin{figure*}[htbp]
    \centering
    \includegraphics[width=1.0\textwidth]{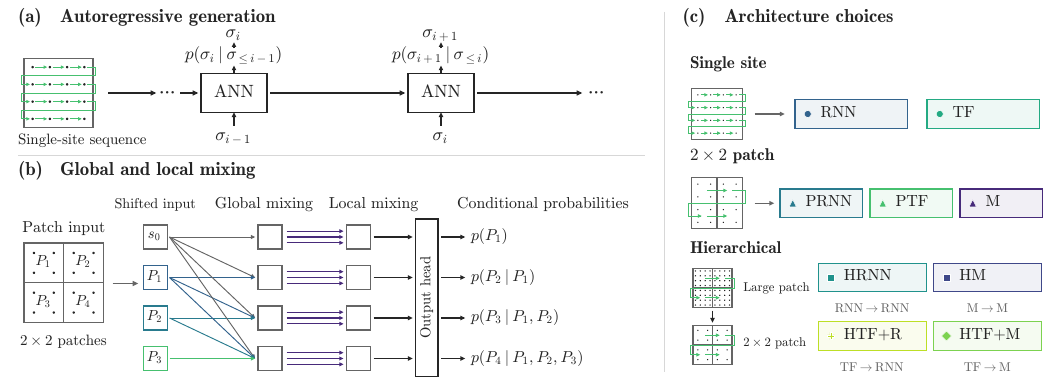}
    \caption{Categorization of ANN architectures used throughout this work. (a) All ANN architectures operate in an autoregressive manner. A sequence of quantum particles is defined in a two-dimensional array using row-major ordering. The state $\sigma_i$ of a particle is used as individual input to the ANN, which outputs a probability distribution $p\left(\sigma_{i+1}\left|\sigma_{\leq i}\right.\right)$ from which the state $\sigma_{i+1}$ of the next particle in the sequence is sampled. The condition on previous sequence elements is naturally encoded in the ANN architectures and can capture correlations in quantum systems. (b) The MLP mixer model uses feed-forward operations to mix information across the input system. Receiving a sequence of patches of $K\times K$ particles as input, it first mixes information globally across all patches. In a subsequent layer, information is mixed locally within each patch. An output layer combines the information and returns probability distributions $\left(P_i\left|P_{\leq i-1}\right.\right)$ over all states of the next patch in the sequence conditioned on all previous sequence elements. See \App{architecture} and \App{B} for a detailed description of the MLP mixer architecture. (c) Definition of all nine ANN architectures used throughout this work. The RNN and transformer model (TF) receive single-particle sequences as input. Patched models, specifically patched RNN (PRNN), patched transformer model (PTF), and MLP mixer model (M), receive input sequences of patches containing $2\times 2$ particles. Hierarchical architectures consist of a main ANN model that receives a sequence of large input patches with $K\times K$ particles and a subsequent ANN model that decomposes each patch into a sequence of $2\times 2$ patches. Here, we consider the hierarchical RNN (HRNN) consisting of two RNNs, the hierarchical MLP mixer architecture (HM) consisting of two MLP mixer models, the hierarchical transformer model with RNN (HTF+R) where the main model is a transformer architecture followed by a subsequent RNN, and the hierarchical transformer architecture with MLP mixer model (HTF+M) that consists of a transformer model followed by an MLP mixer model.}
    \label{fig:architecture_variants}
\end{figure*}

The focus of this work lies in comparing the performance and efficiency of different artificial neural network (ANN) architectures when used as wave function ans\"atze.
Specifically, we consider recurrent neural networks (RNNs), transformer models, and multi-layer perceptron (MLP) mixer models~\cite{tolstikhin2021mlpmixer}.
All three models can efficiently process sequential data in the form $\left(\sigma_1,\dots,\sigma_N\right)$ with sequence length $N$.
By using the network output at sequence step $i$ as input in the next sequence step $i+1$, the ANN architectures can be defined in an autoregressive manner, which allows for sequential data generation~\cite{cho-etal-2014-learning, PixelRNN}.

Such autoregressive ANN models can be used as wave function ans\"atze for quantum many-body systems viewed as sequences of individual correlated particles~\cite{Hibat-a, SharirEtAl, Sprague&Czischek}.
While the sequence definition is clearly defined for one-dimensional physical systems, higher-dimensional systems can be considered by defining a one-dimensional path including all particles~\cite{teoh2024autoregressivemodelpathdependence, RN140}.
In this work, we define a one-dimensional path through two-dimensional physical systems in a row-major order, where we consider particles row-wise from left to right and row by row from top to bottom as illustrated in \Fig{architecture_variants}(a).
The state $\sigma_i$ of a quantum particle then defines the ANN input at sequence step $i$.
The corresponding network output defines a probability distribution,
\begin{align}
    p_{\boldsymbol{\theta}}\left(\sigma_{i+1}\left|\sigma_{0},\dots,\sigma_{i}\right.\right)=p_{\boldsymbol{\theta}}\left(\sigma_{i+1}\left|\sigma_{\leq i}\right.\right),
\end{align}
from which the state $\sigma_{i+1}$ of the next particle is sampled conditioned on the state of previous particles in the sequence.
The index $\boldsymbol{\theta}$ denotes the dependence on the trainable network parameters.
The conditioning on all previous sequence elements is achieved by an intrinsic memory that correlates information over the sequence elements and naturally appears in the definition of the three ANN architectures~\cite{NIPS1999_e6384711, germain2015made}.
In the wave function ansatz, this effect implements correlations among the system particles~\cite{SharirEtAl}.

By running the autoregressive ANN over the entire sequence of particles, we sample quantum state configurations from a joint encoded probability distribution~\cite{SharirEtAl},
\begin{align}
    p_{\boldsymbol{\theta}}\left(\boldsymbol{\sigma}\right)=\prod_{i=1}^Np_{\boldsymbol{\theta}}\left(\sigma_i\left|\sigma_{\leq i-1}\right.\right),
\end{align}
which is interpreted as the squared wave function amplitude,
\begin{align}
    \Psi_{\boldsymbol{\theta}}\left(\boldsymbol{\sigma}\right) &= \left\langle\boldsymbol{\sigma}\left|\Psi_{\boldsymbol{\theta}}\right.\right\rangle
    = \sqrt{p_{\boldsymbol{\theta}}\left(\boldsymbol{\sigma}\right)}.
\end{align}
In the special case of positive, real-valued wave functions, as considered in this work, the squared wave function amplitude provides a complete representation of a quantum state.
For general wave functions, a complex phase term needs to be included in the ANN output~\cite{Hibat-a, HibatAllah2022SupplementingRN}, which is left for future works.
A sample from this joint probability distribution can be interpreted as a quantum state configuration of the entire system and thus emulates a projective measurement outcome in the computational basis that can be used to evaluate operator expectation values~\cite{CarleoandTroyer}.

\subsection{Recurrent Neural Networks}
RNNs are based on a cell consisting of non-linear operations that is applied to each sequential input element to generate an output value~\cite{RN145}.
An intrinsic memory over previous sequence elements is defined by propagating the state of the RNN cell and the included information about individual input values over the sequence.
In the following, we use a gated recurrent unit (GRU) as RNN cell~\cite{cho-etal-2014-properties}.
Since all information about previous sequence elements is encoded in this finite-dimensional cell state, the RNN wave function ansatz is restricted to covering short-range correlations within the input sequence~\cite{Cha_2022, RN134}.
This leads to significant limitations in representing strongly correlated, higher-dimensional quantum systems~\cite{Sprague&Czischek, PhysRevResearch.7.013122}.

\subsection{Transformer Models}
The transformer architecture implements an intrinsic memory over previous sequence elements via a masked self-attention mechanism.
This uses trainable parameters to highlight correlations between individual elements with varying strengths, resulting in output probability distributions that are conditioned on specific sequence elements~\cite{vaswani2017attention}.
The transformer model overcomes the limitations of RNNs and efficiently captures long-range correlations.
This strength comes, however, at a significantly increased computational cost compared to RNNs~\cite{teoh2024autoregressivemodelpathdependence, ayub2026geometryinducedlongrangecorrelationsrecurrent}.

\subsection{Multi-Layer Perceptron Mixer Models}
The MLP mixer model achieves computational efficiency by reducing internal operations to feed-forward vector-matrix multiplications and scalar non-linear transformations.
Combined with reshaping and transposition operations, the architecture can efficiently capture long-range correlations between input elements~\cite{tolstikhin2021mlpmixer}.
Rather than considering a sequence of individual input elements, the MLP mixer model considers a sequence of patches, each consisting of an array of $K\times K$ elements~\cite{tolstikhin2021mlpmixer}.
In such sequences, we distinguish between local information covered within each input patch and global information shared across all patches at individual positions.
In the MLP mixer setup, information is first shared globally via trainable non-linear transformations applied over all input patches.
In a second step, information is similarly shared locally within each patch, where trainable parameters are shared over all global and local transformations~\cite{tolstikhin2021mlpmixer}.
See \Fig{architecture_variants}(b) for an illustration of the information mixing process and \App{architecture} for a detailed description of the MLP mixer architecture.

To become a suitable wave function ansatz, we need to adjust the original MLP mixer architecture.
An autoregressive behavior can be achieved by adding a causal mask in the global information mixing layer, eliminating information from input elements that appear later in the sequence.
We further apply a softmax activation function to the network output, so that the output at each sequence step can be interpreted as the probability distribution over all possible states of the next patch of $K\times K$ particles conditioned on the states of previous patches in the sequence.
We discuss the modified MLP mixer architecture in detail in \App{B}.

\subsection{Patched and Hierarchical Architectures}
In consistency with the MLP mixer model, which naturally considers patches of $K\times K$ input elements, we further consider patched RNN and transformer models.
These receive the state of a two-dimensional patch of $K\times K$ particles as input and return the probability distribution over all possible states of the next patch conditioned on all previous sequence patches.
Such patched setups significantly reduce the computational runtime and improve the representational accuracy compared to single-particle sequences~\cite{Sprague&Czischek, Luciano}.

However, the number of possible states in a patch of particles scales exponentially with the patch size, leading to an exponentially growing network output dimension.
The resulting exponentially scaling computational cost restricts our considerations to input patches of $2\times 2$ particles.
To allow for an efficient usage of larger input patches, we consider hierarchical extensions consisting of a main ANN model that receives large input patches with $K>2$ and feeds its output into a subsequent ANN model that hierarchically decomposes the large input patch into a sequence of smaller patches of $2\times 2$ particles.
This setup is similar to large, patched transformer models~\cite{Sprague&Czischek} and leads to a fixed output dimension.

In the following, we compare the performance of nine ANN architectures when used as wave function ans\"atze, namely the RNN and patched RNN, the transformer architecture and patched transformer architecture, the MLP mixer architecture, a hierarchical model consisting of two RNNs, a hierarchical model consisting of two MLP mixer architectures, and two hierarchical models consisting of a transformer architecture followed by an RNN or an MLP mixer architecture, respectively.
The characteristics of the nine model architectures as well as their nomenclature is summarized in \Fig{architecture_variants}(c) and \Tab{architecture-plot-key} in \App{plots}.
To achieve a fair performance comparison for the considered ANN architectures, we fix the number of variational parameters within the same order of magnitude for all models--$\approx 5{,}000$ parameters for lattices of size $4 \times 4$, $\approx 10{,}000$ parameters for lattices of size $8 \times 8$, $\approx 20{,}000$ parameters for lattices of size $16 \times 16$, and $\approx 30{,}000$ parameters for lattices of size $24 \times 24$ (see \Tab{full-flops} in \App{computational-cost} for more details).

\section{Physical Systems}
We compare the accuracy and computational efficiency of the different ANN architectures in finding ground state representations for the two-dimensional transverse-field Ising model and a two-dimensional array of Rydberg atoms.
The Hamiltonians underlying both systems are stoquastic and have positive, real-valued ground state wave functions~\cite{Hibat-a, Sprague&Czischek}. 
For both systems, we study square lattices with $4\times 4$ and $8\times 8$ particles and open boundary conditions.
For the Rydberg atom arrays, we further consider systems with $16\times 16$ and $24\times 24$ atoms.

To find ground state representations, we train the ANN models variationally, tuning the weight and bias parameters to minimize the energy expectation value $\left\langle{E}\right\rangle$ for a given Hamiltonian $\hat{\mathcal{H}}$ as evaluated on $N_s$ samples $\boldsymbol{\sigma}_n$ drawn from the generative ANN~\cite{Hibat-a},
\begin{align}
    \left\langle E\right\rangle &= \frac{1}{N_s}\sum_{n=1}^{N_s}E_{\mathrm{loc}}\left(\boldsymbol{\sigma}_n\right),
\end{align}
with local energy, 

\begin{align}
E_{\mathrm{loc}}\left(\boldsymbol{\sigma}\right) &= \sum_{\boldsymbol{\sigma}^\prime} \left\langle\boldsymbol{\sigma}\left|\hat{\mathcal{H}}\right|\boldsymbol{\sigma}^\prime\right\rangle\frac{\psi_{\boldsymbol{\theta}}\left(\boldsymbol{\sigma}^\prime\right)}{\psi_{\boldsymbol{\theta}}\left(\boldsymbol{\sigma}\right)}.
\end{align}

\subsection{Two-Dimensional Transverse Field Ising Model}
We first consider ground states of the two-dimensional transverse-field Ising model (TFIM).
This model for $N=N_x\times N_y$ qubits is described by the Hamiltonian
\begin{align}
    \hat{\mathcal{H}}_{\mathrm{TFIM}} &= -J\sum_{\left\langle i,j\right\rangle}\hat{\sigma}_i^z\hat{\sigma}_j^z-h\sum_{i=1}^N\hat{\sigma}_i^x,
\end{align}
where $\hat{\sigma}_i^{\alpha}$ describes the $\alpha$-Pauli matrix ($\alpha\in\left\{x,y,z\right\}$) acting on site $i$ and the first sum runs over all nearest-neighbor qubit pairs $\left\langle i,j\right\rangle$.
The nearest-neighbor coupling strength is defined by $J$, which we fix to $J=1$ in the following.
The transverse field $h$ then defines a second-order quantum phase transition at $h_{\mathrm{c}}\approx 3.044$ between a ferromagnetic phase ($h<h_{\mathrm{c}}$) and a paramagnetic phase ($h>h_{\mathrm{c}}$).

\begin{figure}[htbp]
    \centering
    \input{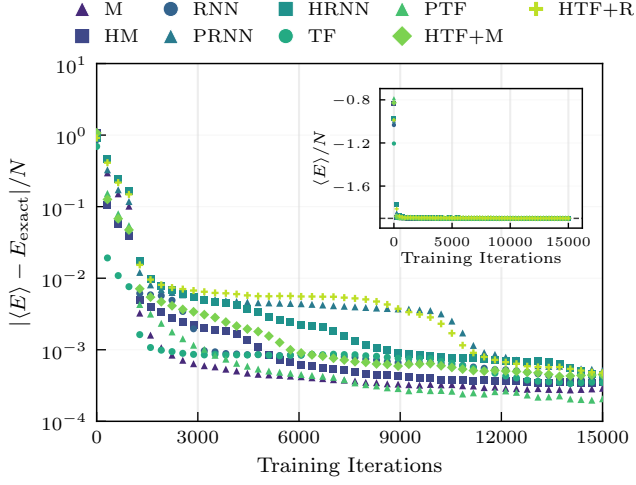}
    \caption{Search for TFIM ground states in an $8\times 8$-site lattice at $h = 1.0$. The nine different ANN architectures (different colors) are variationally trained to represent the ground state. The main plot shows the evolution of the absolute energy-density error $\left|\left\langle E\right\rangle - E_{\mathrm{exact}}\right|/N$ for all ANN architectures during training for $15{,}000$ iterations where we compare to the exact ground state energy density obtained via tensor network simulations~\cite{itensor}. The inset shows the corresponding energy density $\left\langle E\right\rangle /N$, where all data points start to overlap quickly with the exact energy indicated by a black horizontal line. Data points correspond to averages of the preceding $1000$ training iterations. Hierarchical ANN architectures use initial input patches of $4\times 4$ sites. All ANN architectures successfully converge to the ground state energy.}
    \label{fig:2}
\end{figure}

\Fig{2} shows the behavior of the energy expectation value during the training process for all nine ANN architectures on an $8\times 8$ lattice in the ferromagnetic phase, $h=1$.
The main plot shows the absolute energy-density error, $\left|\left\langle E\right\rangle-E_{\mathrm{exact}}\right|/N$, as a function of the training iterations, where we obtain the exact energy using tensor network simulations~\cite{itensor}.
The inset shows the corresponding energy as a function of training iterations.
In both cases, results from the nine considered ANN architectures are shown in different colors. 
The definition of the individual network architectures is provided in \Fig{architecture_variants} and summarized in \Tab{architecture-plot-key} in \App{plots}.
All architectures converge to accurate ground state representations in about $10{,}000$ training iterations. 
However, we find that the patched transformer model, the MLP mixer model, and all hierarchical models recover the ground state significantly faster, in less than $5{,}000$ training iterations. 

\subsection{Rydberg Atom Arrays}
As a second benchmark, we consider Rydberg atoms arranged on square lattices with $N=N_x\times N_y$ sites.
The system is described by the Rydberg Hamiltonian~\cite{Hibat-b, Sprague&Czischek},
\begin{align}
    \hat{\mathcal{H}}_{\mathrm{Rydberg}} &= -\frac{\Omega}{2}\sum_{i=1}^{N}\hat{\sigma}_i^x-\delta\sum_{i=1}^N\hat{n}_i+\sum_{i,j=1}^{N}V_{i,j}\hat{n}_i\hat{n}_j,
\end{align}
with off-diagonal operator $\hat{\sigma}_{i}=\left|g_i\right\rangle\left\langle r_i\right|+\left|r_i\right\rangle\left\langle g_i\right|$ driving transitions between ground states $\left|g_i\right\rangle$ and highly excited Rydberg states $\left|r_i\right\rangle$, and occupation number operator $\hat{n}_i=\left|r_i\right\rangle\left\langle r_i\right|$.
Here, $\delta$ and $\Omega$ describe the detuning between ground and Rydberg levels and the frequency of Rabi oscillations between the two states, respectively.
In experimental realizations, both effects are created via external laser driving.
Atoms at positions $\boldsymbol{r}_i$ and $\boldsymbol{r}_j$ interact via long-range van-der-Waals interactions as described by $V_{i,j}=\Omega R_{\mathrm{b}}^6/\left|\boldsymbol{r}_i-\boldsymbol{r}_j\right|^6$, where any excitation is energetically penalized within the Rydberg blockade radius $R_{\mathrm{b}}$ around an excited atom~\cite{RydbergBlockade}.
We have set the lattice spacing to $a=1$.
By tuning the Hamiltonian parameters, the system can be prepared in a variety of phases of matter separated by different kinds of phase transitions~\cite{Sprague&Czischek, PhaseTransition}.
In the following, we fix $\Omega=1$ and $R_{\mathrm{b}}=3^{1/6}$, while we consider choices for $\delta$ that prepare the system in different phases.

\begin{figure}[htbp]
    \centering
    \input{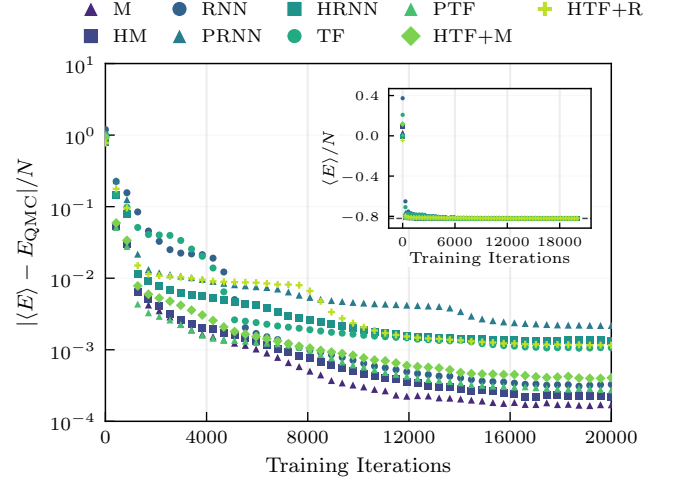}
    \caption{Search for ground states of $8\times 8$ Rydberg atom arrays in the checkerboard phase at $\delta=2$. The training behavior over $20{,}000$ iterations is benchmarked on energy values obtained via quantum Monte Carlo simulations~\cite{merali_stochastic_2021} for the nine considered ANN architectures (different colors). The main plot shows the absolute energy-density error $\left|\left\langle E\right\rangle-E_{\mathrm{QMC}}\right|/N$ as a function of the training iterations, while the inset shows the corresponding energy density values $\left\langle E\right\rangle/N$. Data points correspond to averages over the $1000$ preceding training iterations. Hierarchical ANN architectures use initial input patches of $4\times 4$ atoms. All ANN architectures quickly converge to accurate ground state representations, leading to a quick overlap of all data points with the quantum Monte Carlo energy values (black dashed line) in the inset.}
    \label{fig:3}
\end{figure}

\Fig{3} shows the performance of the different ANN architectures for finding ground state representations of an $N=8\times 8$ Rydberg atom array on a square lattice in the checkerboard phase at $\delta=2$. 
We benchmark the nine ANN architectures by comparing to energy expectation values $E_{\mathrm{QMC}}$ obtained with quantum Monte Carlo simulations~\cite{merali_stochastic_2021} that we consider as accurate estimate of exact energy expectation values.
The main plot shows the absolute energy-density error $\left|\left\langle E\right\rangle- E_{\mathrm{QMC}}\right|/N$ as a function of the training iterations, while the inset shows the evolution of the corresponding expected energy density as function of the variational training iterations.
All network architectures find accurate ground state representations within $12{,}000$ training iterations.
The patched and hierarchical models show significantly faster convergence than the RNN and the transformer architecture with single-atom inputs.
This behavior matches the observations for the TFIM in \Fig{2} and can be understood from the fact that grouping individual sites into patches treats localized regions as atomic units.
This way, local correlations inside the patches are captured more accurately in the ANN-based representation.
The MLP mixer architecture naturally uses such a patch-based encoding to combine local and non-local information, making it well suited for representing lattice-based quantum systems~\cite{tolstikhin2021mlpmixer}.

\section{Results}

Having observed in \Fig{2} and \Fig{3} that all investigated ANN models do find representations of ground states in the TFIM and the Rydberg atom array whose accuracies meet the state of the art~\cite{Carrasquilla2026} for one specific set of Hamiltonian parameters, we now focus on comparing the performance and computational efficiency of these models in different phases of matter and for different physical system sizes.
For each configuration, we evaluate the number of floating-point operations (FLOPs) in a single network pass as explained in \App{computational-cost} and aggregate the absolute ground state energy error after $15{,}000$ training iterations for TFIM simulations, after $20{,}000$ training iterations for Rydberg array simulations with $N=4\times 4$, $N=8\times 8$, and $N=16\times 16$ atoms, and after $25{,}000$ training iterations for Rydberg array simulations with $N=24\times 24$ atoms by averaging over three points in the respective phase diagram.
We provide the energy training curves for all system configurations in \Fig{all_tfim_plots} and \Fig{all_rydberg_plots} in \App{plots}, demonstrating that all training algorithms have converged after the corresponding number of training iterations.
Considering the number of floating-point operations provides a hardware-agnostic measure of the computational costs.

In the two-dimensional TFIM, we consider points in the ferromagnetic phase ($h=1$), in the paramagnetic phase ($h=5$), and in the vicinity of the phase transition ($h=3.044$).
In the Rydberg atom array, we similarly consider points in the disordered phase ($\delta=0.4$), in the checkerboard phase ($\delta=2$), and in the vicinity of the transition between the two ($\delta=1.2$).
Averaging the performance over different regimes in the phase diagram allows us to compare the computational cost and representational accuracy of all ANN architectures in a general manner. 
Each data point is further averaged over five independently initialized training runs.

\begin{figure}[htbp]
    \centering
    \includegraphics[width=1.0\columnwidth]{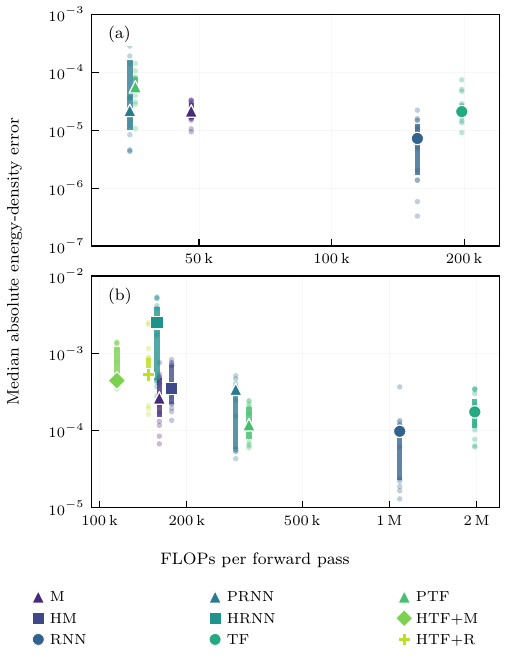}
    \caption{Median absolute energy-density error as a function of compute resources for the two-dimensional TFIM Hamiltonian. We evaluate the absolute energy-density error $\left|\left\langle E\right\rangle-E_{\mathrm{exact}}\right|/N$ at three points in the phase diagram ($h=1$, $h=3.044$, $h=5$) after $15{,}000$ training iterations for the different network architectures. Each data point shows the median of the energy density error evaluated over the three points in the phase diagram and five independent training runs as a function of the computational efficiency evaluated as FLOPs in a single forward pass. Panel (a) shows results for a system of $4\times 4$ sites, while panel (b) considers a system of $8\times 8$ sites. Solid data points denote the median absolute energy-density error, solid error bars correspond to the interquartile range, and shaded data points show results from all $15$ individual runs. While the patched RNN and patched transformer model show the most promising performance for the smaller system, hierarchical architectures and MLP mixer models appear stronger in the larger system size.}
    \label{fig:4}
\end{figure}

We start with investigating the ANN model performances for the two-dimensional TFIM.
The results are shown in \Fig{4}, where we plot the average absolute energy-density error after $15{,}000$ training iterations versus the number of FLOPs used in a single forward-pass in each ANN architecture.
We consider systems with $N=4\times 4$ qubits in panel (a) and $N=8\times 8$ qubits in panel (b).
The hierarchical ANN architectures are only considered in panel (b) where they receive initial input patches of $4\times 4$ qubits.
In \Fig{4}, the best performing ANN architecture can be found close to the origin of the plots where the error and the number of FLOPs are minimized.
All ANN architectures reach satisfying accuracy for the TFIM, so that the focus lies in comparing the computational efficiency.
In these means, for $N=4\times 4$ qubits, the patched transformer architecture and the patched RNN show the best performance, while the MLP mixer requires more FLOPs. 
The single-qubit RNN and transformer model require significantly more FLOPs, which is explained by the longer sequence lengths.
For $N=8\times 8$ qubits, the hierarchical models outperform all patched and single-qubit models.
Remarkably, the hierarchical MLP mixer model requires a comparable amount of FLOPs to the MLP mixer model and the hierarchical transformer architectures while achieving a competitive representational accuracy. 
Combining the hierarchical transformer model with the MLP mixer model requires the least amount of FLOPs.
Those results affirm the expected low computational cost and high representational accuracy of the MLP mixer model.

\begin{figure*}[htbp]
    \centering
    \includegraphics[width=1.0\textwidth]{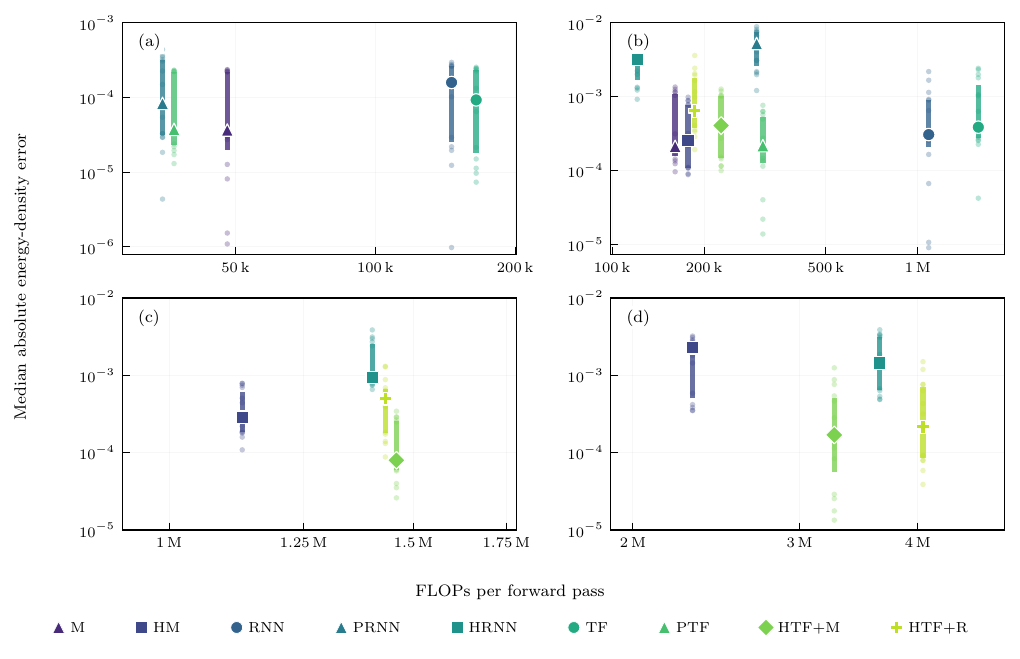}
    \caption{Median absolute energy-density error as a function of compute resources for Rydberg atom arrays. We compare the median of the absolute energy-density error $\left|\left\langle E\right\rangle-E_{\mathrm{QMC}}\right|/N$ after $20{,}000$ training iterations for $N=4\times 4$, $N=8\times 8$, and $N=16\times 16$ and after $25{,}000$ training iterations for $N=24\times 24$, evaluated over three points in the phase diagram ($\delta=0.4$, $\delta=1.2$, $\delta=2$) for the different considered ANN architectures (different colors). Each median is furthermore calculated over five independent training runs and plotted as a function of the required FLOPs per forward pass. Individual panels show results for arrays of $4\times 4$ atoms [panel (a)], $8\times 8$ atoms [panel (b)], $16\times 16$ atoms [panel (c)], and $24\times 24$ atoms [panel (d)]. Solid data points show the median absolute energy-density error, solid error bars correspond to the interquartile range, and shaded data points indicate results from the individual runs. The computational efficiency of the MLP mixer model is clearly visible for large system sizes, while its representational accuracy decreases. Overall, the hierarchical transformer model combined with the MLP mixer architecture shows the most promising performance at large scales.}
    \label{fig:5}
\end{figure*}

\begin{figure}
    \centering
    \includegraphics[width=1.0\columnwidth]{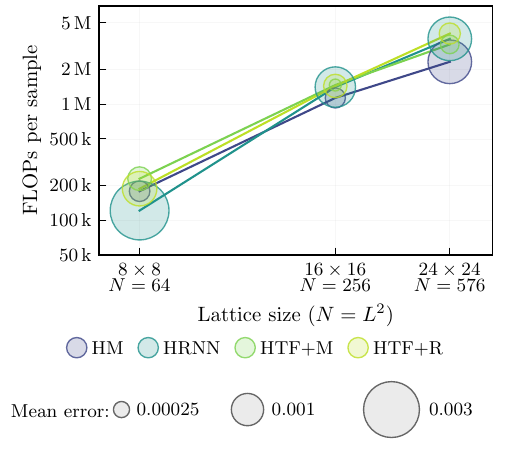}
    \caption{Scaling of computational cost for large Rydberg atom arrays. We compare the number of FLOPs required by the four hierarchical ANN architectures (different colors) for a single forward pass as function of the physical system size $N=L\times L$ in a two-dimensional Rydberg atom array. Data points show the number of FLOPs as function of the system size, where lines connect the points as a guide to the eye. Circles around the data points indicate the error in energy density after $20{,}000$ training iterations ($25{,}000$ iterations for $N=24\times 24$) averaged over three points in the phase diagram ($\delta=0.4$, $\delta=1.2$, $\delta=2$) with five independent runs each. We use quantum Monte Carlo simulations~\cite{merali_stochastic_2021} to estimate the exact energy. The shown data is extracted from the same simulations as in \Fig{5}. While the hierarchical MLP mixer model shows the most promising scaling behavior, its error increases significantly for large system sizes. The hierarchical transformer architecture with MLP mixer model shows the most promising behavior in terms of scaling of FLOPs and representational accuracy.}
    \label{fig:6}
\end{figure}

We perform a similar analysis for the Rydberg atom array in \Fig{5}.
Here, we consider systems with $N=4\times 4$ atoms [panel (a)], $N=8\times 8$ atoms [panel (b)], $N=16\times 16$ atoms [panel (c)], and $N=24\times 24$ atoms [panel (d)].
Similarly to the TFIM investigations, we only consider hierarchical architectures for $N=8\times 8$, $N=16\times 16$, and $N=24\times 24$ atoms.
Furthermore, the systems with $N=16\times 16$ and $N=24\times 24$ atoms no longer allow for the consideration of single-atom and patched models due to exceeding computational costs.
Our observations in panels (a) and (b) agree with what we observed for the TFIM in \Fig{4}.
For small systems ($N=4\times 4$ atoms), the patched RNN shows the most promising performance followed by the patched transformer architecture.
The MLP mixer model achieves similar performance but requires a larger number of FLOPs.
For intermediate system sizes ($N=8\times 8$ atoms), the hierarchical MLP mixer and MLP mixer architectures achieve the best performance by showing the highest computational efficiency. 
The hierarchical RNN requires less FLOPs but shows an order-of-magnitude larger error.
Looking at the large-scale systems ($N=16\times 16$ and $N=24\times 24$ atoms), we find the lowest errors in the hierarchical transformer architecture combined with the MLP mixer model.
However, these models require significantly more FLOPs than the hierarchical MLP mixer model, which shows a satisfying representation accuracy for $N=16\times 16$ atoms.
For $N=24\times 24$ atoms, the hierarchical MLP mixer architecture also requires the smallest number of FLOPs but only achieves a representational accuracy comparable to the hierarchical RNN while transformer-based hierarchical models show significantly smaller errors.
For large system sizes, combining the hierarchical transformer architecture with the MLP mixer model thus shows the most promising combination of representational accuracy and FLOP requirement.

In \Fig{6}, we analyze the scaling of computational cost with increasing system size in the two-dimensional Rydberg atom array.
We only consider hierarchical ANN architectures that allow efficient consideration of large-scale systems with up to $N=24\times 24$ atoms.
While the data is obtained from the same simulations as in \Fig{5}, this arrangement clearly shows that the hierarchical RNN suffers the highest energy error while the hierarchical MLP mixer model achieves the most promising FLOP scaling behavior.
However, at $N=24\times 24$, the hierarchical MLP mixer shows an energy error of similar size to the hierarchical RNN.
While the consideration of larger system sizes is necessary to draw a certain conclusion, our simulations indicate that the most promising choice in terms of FLOPs and representational accuracy for investigating large-scale quantum systems is the hierarchical transformer architecture with MLP mixer model.

\section{Discussion and Outlook}
In this work, we demonstrate that the multi-layer perceptron (MLP) mixer model as a fully feed-forward, attention-free artificial neural network architecture can serve as an efficient variational wave function ansatz to accurately represent ground states of higher-dimensional quantum many-body systems.
We show that this model can successfully find ground state representations in various phases of matter while meeting accuracy standards of state-of-the-art variational ans\"{a}tze~\cite{Carrasquilla2026}.
We further introduce hierarchical artificial neural network (ANN) architectures that receive large input patches of multiple qubits and decompose them hierarchically into smaller patches. 
We construct those network architectures by combining recurrent neural networks (RNNs), transformer models, and the MLP mixer model.
When comparing computational efficiencies of different ANN architectures, we find that the hierarchical transformer architecture combined with the MLP mixer model outperforms state-of-the-art methods such as (patched) transformer models and RNNs when considering system sizes beyond $8\times 8$ qubits.
However, for smaller system sizes, the patched RNN shows the best combination of representational accuracy and computational efficiency. 

Our observations can be justified by considering the computational performance of the different ANN architectures.
The computational cost of the RNN scales linearly with the considered sequence length~\cite{Hibat-a} , while the transformer model shows quadratic scaling with the input sequence length~\cite{Sprague&Czischek, rende2026scalinglawsneuralnetworkquantum}.
Patch-based models, such as the patched RNN and the patched transformer model, reduce the effective sequence length compared to single-qubit considerations, leading to faster run times.
At the same time, patching the input system improves the encoding of local correlations that can be directly integrated in the network setup ~\cite{trockman2022patches, Sprague&Czischek, s3kd-bc4q}.
The MLP mixer model shows linear scaling with input sequence length while being naturally patch-based~\cite{tolstikhin2021mlpmixer}.
Our results show furthermore that single-level patching remains insufficient at large system sizes.
The hierarchical constructions outperform the patched models by introducing a decomposition at multiple scales.
While the main network model captures the global structure of large input patches using a coarse-grained representation, the inner model resolves local correlations conditioned on this global context.

Our study thus shows that the best choice of ANN architecture for ground state representations depends strongly on the system under consideration.
However, using the hierarchical network architecture consisting of a transformer model and an MLP mixer model as variational wave function ansatz allows for ground state representations of large-scale quantum systems beyond limitations of conventional ANN-based wave function ans\"atze.
With this, it allows for detailed investigations of phase diagrams at large scales, promising insights into novel quantum phases of matter and quantum phase transitions.
While we focus on systems with even numbers of particles in each direction, our introduced hierarchical models can similarly be used for odd-dimensional system sizes by adjusting the patch sizes accordingly, making our models generally applicable.
The fully feed-forward nature of the MLP mixer model further suggests straight-forward implementations on specialized hardware such as spiking neuromorphic devices, promising additional computational improvements.

\section*{Acknowledgments}
We thank Sehmimul Hoque and the APRIQuOt team for fruitful insights and discussions.
We acknowledge the support of the Natural Sciences and Engineering Research Council of Canada (NSERC), NSERC CREATE/ 575280-2023 - Training in Materials for Quantum Technologies (MaQTech).
This research was enabled in part by computational resources and support provided by the BC DRI Group (bcdri.ca) and the Digital Research Alliance of Canada (alliancecan.ca).
Additional computational resources were provided by Google Colaboratory (colab.research.google.com) and Lambda (lambdalabs.com).

\section*{Code Availability}
We wrote our codes in the Python programming language. We used the PyTorch (v2.12) ~\cite{cite_pytorch} deep learning framework, jaxtyping~\cite{cite_jaxtyping} and tqdm~\cite{cite_tqdm} in designing and training all models. NumPy~\cite{cite_numpy}, Matplotlib~\cite{cite_matplotlib} and Polars~\cite{cite_polars} were used to analyze the data and generate plots. 
Simulations were run on NVIDIA H100 instances [provided by the BC DRI Group (bcdri.ca), the Digital Research Alliance of Canada (alliancecan.ca), and Lambda (lambdalabs.com)], NVIDIA RTX PRO A6000 instances [provided by Google Colaboratory (colab.research.google.com)], and a private NVIDIA L40 instance.
The data presented in this paper and the code are available on GitHub~\cite{code_repo}.

\appendix
\section{The MLP Mixer Architecture}
\label{app:architecture}

The multi-layer perceptron (MLP) mixer architecture relies on MLPs, reshaping, transposing, and scalar nonlinearities~\cite{tolstikhin2021mlpmixer}.
Its architecture can be considered consisting of two operations, both implemented with MLPs.
The first operation mixes information per location in the input across multiple channels, while the second operation mixes information across all locations in each input channel.
We consider the two operations as global and local information mixing, respectively, which allows for covering long- and short-range correlations in the input.

The MLP mixer architecture receives a two-dimensional image $X^{\left(\mathrm{in}\right)}$ of shape $H\times W$, which is broken into $S=\left(H\times W\right)/P^2$ non-overlapping patches of size $P\times P$.
Each patch is flattened and individually sent to a fully-connected ANN layer, projecting it to a higher dimension $C$.
With this, the input is represented in a data array $X^{\left(1\right)}$ of dimension $S\times C$, where we refer to $S$ as the number of tokens (patches) and to $C$ as the number of channels (projection features).
These dimensions will remain constant throughout the rest of the network operations.
All patches share the same dimension $D_S=P\times P$ and all channels share the same dimension $D_C$.
These dimensions are hyper-parameters that are initially fixed.
The patch dimension $D_S$ defines the number of input patches $S$, while $D_C$ is independent of the patch size.
With this, the computational complexity of the MLP mixer architecture is linear in the number of input patches $S$ and the patch size $D_S$, making it a computationally efficient model with potential to scale to large input sizes.

First, the information is mixed globally via a fully connected layer that receives all tokens in a single channel and provides an output of the same dimension,
\begin{align}
    \boldsymbol{X}^{\left(2\right)}_{*,i} &= W^{\left(1,2\right)}\boldsymbol{X}_{*,i}^{\left(1\right)}.
\end{align}
Here, the index $i\in\left\{1,\dots,C\right\}$ denotes the individual channels, $W^{\left(1,2\right)}$ is a weight matrix of dimension $S\times S$, and the index $*$ indicates that we consider the individual columns of $X^{\left(1\right)}$ as input vectors, while the output vectors can be interpreted as columns of an output matrix $X^{\left(2\right)}$.
The weight matrix $W^{\left(1,2\right)}$ is shared over all channels.

In the second step, the information is mixed locally via a fully connected layer acting on all channels for each token,
\begin{align}
    \boldsymbol{X}^{\left(3\right)}_{j,*} &= W^{\left(2, 3\right)}\boldsymbol{X}_{j,*}^{\left(2\right)},
\end{align}
with the index $j\in\left\{1,\dots,S\right\}$ denoting individual tokens.
Here, $W^{\left(2,3\right)}$ is a weight matrix of dimension $C\times C$ that is shared over all tokens and the index $*$ shows that we now consider individual rows of $X^{\left(2\right)}$ as input vectors.
Each output vector is similarly interpreted as a row of the output matrix $X^{\left(3\right)}$.

Each mixing operation is followed by a non-linear activation function, where a Gaussian Error Linear Unit (GELU) function is used~\cite{GELU}, and another linear transformation with trainable weight parameters $V^{\left(1,2\right)}$ and $V^{\left(2,3\right)}$ for global and local mixing operations, respectively.
Furthermore, each mixing operation is preceded by a layer normalization and succeeded by a skip connection, such that the full information flow in the MLP mixer architecture is given by
\begin{align}
    \boldsymbol{X}^{\left(2\right)}_{\star,i} &= \boldsymbol{X}_{*,i}^{\left(1\right)}+V^{\left(1,2\right)}\mathrm{GELU}\left[W^{\left(1,2\right)}\tilde{\boldsymbol{X}}_{*,i}^{\left(1\right)}\right],\\
    \boldsymbol{X}_{j,*}^{\left(3\right)} &= \boldsymbol{X}_{j,*}^{\left(2\right)}+V^{\left(2,3\right)}\mathrm{GELU}\left[W^{\left(2,3\right)}\tilde{\boldsymbol{X}}_{j,*}^{\left(2\right)}\right].
\end{align}
Here, $V^{\left(i,j\right)}$ denote weight matrices of the same dimensions as $W^{\left(i,j\right)}$ that are shared similarly over all channels and tokens, respectively.
The notation $\tilde{X}$ denotes that a layer normalization has been applied to the matrix $X$.
The original MLP mixer architecture has been applied for image classification, so that a global average pooling layer across the patch dimension followed by a classifier head are applied to $X^{\left(3\right)}$ to obtain the desired output~\cite{tolstikhin2021mlpmixer}.

Other MLP-based mixer models share the global and local mixing framework, but differ in the details of the network architecture~\cite{LIUParadigmShift}.

\section{MLP Mixer Architecture as Wave Function Ansatz}
\label{app:B}

\begin{figure}[h]
\includegraphics[width=8cm]{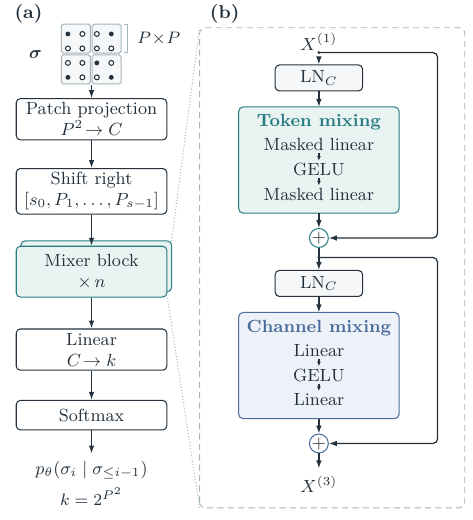}
\caption{Autoregressive MLP mixer architecture, where arrowheads indicate the direction of information flow. \textbf{(a)} The two-dimensional quantum system passes through a patching layer where it is divided into $S$ non-overlapping regions consisting of $P\times P$ particles. Each patch is embedded in a vector of dimension $C$ and a right-shift operation is applied. The following mixing layers can be stacked. The output of the mixing operations is propagated through a linear layer that reduces to the output dimension $k=2^{P\times P}$ to encode a probability distribution over all possible states of the next patch in the sequence. A softmax activation function is used to obtain an output that can be interpreted as the conditional probability distribution from which the next patch can be sampled. \textbf{(b)} A pre-layer normalization $\mathrm{LN}_C$ is applied across the channel dimension before each mixing operation. In the token mixing layer, a causal mask is applied to both linear operations. Each mixing layer consists of a linear layer followed by a GELU activation function, and another linear layer. Furthermore, each mixing layer is succeeded by a skip connection.}
\label{modified mixer}
\centering
\end{figure}

Using the MLP mixer architecture as a wave function ansatz requires tailored modifications of the original architecture as introduced in~\cite{tolstikhin2021mlpmixer}.
In order to obtain a network output that can be interpreted as a probability distribution over the states of the next sequence element, we apply a softmax activation function to the output $X^{\left(3\right)}$ of the local mixing layer, rather than a global pooling operation and a classifier head as in~\cite{tolstikhin2021mlpmixer}.
As we only consider ground states of stoquastic Hamiltonians in this work, this output is sufficient to be interpreted as the amplitude of the real-valued wave function. 
In more general cases, an additional output head can be added to include information about complex phases, similar to the RNN wave function ansatz~\cite{Hibat-a}. 

Next, we implement the autoregressive setup in the MLP mixer architecture, which is not provided in the original setup~\cite{tolstikhin2021mlpmixer}.
For this, the architecture needs to be autoregressive in the input tokens that correspond to patches of quantum particles.
Similarly to autoregressive implementations of transformer architectures~\cite{MultipurposeTransformer}, we apply a causal mask before information is shared globally across inputs.
For the MLP mixer architecture to be truly causal, this causal mask must be applied in both linear layers appearing in the global mixing layer.
Here, we use MADE-style (Masked Autoencoder for Density Estimation) connectivity to implement the masking~\cite{MADE} and ensure that only information from patches that appear earlier in the input sequence contributes to the mixing.

The skip connection that appears after the token-mixing operation naturally breaks the order of the input sequence as information from the current sequence element can leak in.
To resolve this issue, we apply a right-shift operation that drops the embedding of the current sequence element $P_i$ and prepends a learned start embedding.
The input sequence then becomes,
\begin{align}
    \left[P_1,P_2,\dots,P_i\right] \rightarrow \left[s_0,P_1,P_2,\dots,P_{i-1}\right].
\end{align}
This way, the residual stream does not carry information about the currently considered patch $P_i$.
In the normalization operations, we further only normalize over the channel dimension and not over the token (input) dimension to not break the autoregressive ordering.

\section{Training Performances}
\label{app:plots}

\begin{table*}[t]
  \setlength{\columnwidth}{\textwidth}
  \setlength{\linewidth}{\textwidth}
  \setlength{\hsize}{\textwidth}
  \centering
  \caption{Architecture key for the training plots. Symbol colors, marker shapes, and network nomenclature are identical in all panels in \Fig{2}, \Fig{3}, \Fig{4}, \Fig{5}, \Fig{6}, \Fig{all_tfim_plots}, and \Fig{all_rydberg_plots}.}
  \label{tab:architecture-plot-key}
  \begin{ruledtabular}
\begin{tabular}{cll}
Symbol & Label & Architecture \\
\hline
\raisebox{-0.25em}{
\begingroup%
\makeatletter%
\begin{pgfpicture}%
\pgfpathrectangle{\pgfpointorigin}{\pgfqpoint{0.220000in}{0.140000in}}%
\pgfusepath{use as bounding box, clip}%
\begin{pgfscope}%
\pgfsetbuttcap%
\pgfsetmiterjoin%
\definecolor{currentfill}{rgb}{1.000000,1.000000,1.000000}%
\pgfsetfillcolor{currentfill}%
\pgfsetlinewidth{0.000000pt}%
\definecolor{currentstroke}{rgb}{1.000000,1.000000,1.000000}%
\pgfsetstrokecolor{currentstroke}%
\pgfsetdash{}{0pt}%
\pgfpathmoveto{\pgfqpoint{0.000000in}{0.000000in}}%
\pgfpathlineto{\pgfqpoint{0.220000in}{0.000000in}}%
\pgfpathlineto{\pgfqpoint{0.220000in}{0.140000in}}%
\pgfpathlineto{\pgfqpoint{0.000000in}{0.140000in}}%
\pgfpathlineto{\pgfqpoint{0.000000in}{0.000000in}}%
\pgfpathclose%
\pgfusepath{fill}%
\end{pgfscope}%
\begin{pgfscope}%
\pgfpathrectangle{\pgfqpoint{0.000000in}{0.000000in}}{\pgfqpoint{0.220000in}{0.140000in}}%
\pgfusepath{clip}%
\pgfsetbuttcap%
\pgfsetroundjoin%
\definecolor{currentfill}{rgb}{0.280255,0.165693,0.476498}%
\pgfsetfillcolor{currentfill}%
\pgfsetlinewidth{0.351312pt}%
\definecolor{currentstroke}{rgb}{1.000000,1.000000,1.000000}%
\pgfsetstrokecolor{currentstroke}%
\pgfsetdash{}{0pt}%
\pgfsys@defobject{currentmarker}{\pgfqpoint{-0.036747in}{-0.036747in}}{\pgfqpoint{0.036747in}{0.036747in}}{%
\pgfpathmoveto{\pgfqpoint{0.000000in}{0.036747in}}%
\pgfpathlineto{\pgfqpoint{-0.036747in}{-0.036747in}}%
\pgfpathlineto{\pgfqpoint{0.036747in}{-0.036747in}}%
\pgfpathlineto{\pgfqpoint{0.000000in}{0.036747in}}%
\pgfpathclose%
\pgfusepath{stroke,fill}%
}%
\begin{pgfscope}%
\pgfsys@transformshift{0.110000in}{0.070000in}%
\pgfsys@useobject{currentmarker}{}%
\end{pgfscope}%
\end{pgfscope}%
\end{pgfpicture}%
\makeatother%
\endgroup%
} & $\mathrm{M}$ & Patched MLP mixer \\
\raisebox{-0.25em}{
\begingroup%
\makeatletter%
\begin{pgfpicture}%
\pgfpathrectangle{\pgfpointorigin}{\pgfqpoint{0.220000in}{0.140000in}}%
\pgfusepath{use as bounding box, clip}%
\begin{pgfscope}%
\pgfsetbuttcap%
\pgfsetmiterjoin%
\definecolor{currentfill}{rgb}{1.000000,1.000000,1.000000}%
\pgfsetfillcolor{currentfill}%
\pgfsetlinewidth{0.000000pt}%
\definecolor{currentstroke}{rgb}{1.000000,1.000000,1.000000}%
\pgfsetstrokecolor{currentstroke}%
\pgfsetdash{}{0pt}%
\pgfpathmoveto{\pgfqpoint{0.000000in}{0.000000in}}%
\pgfpathlineto{\pgfqpoint{0.220000in}{0.000000in}}%
\pgfpathlineto{\pgfqpoint{0.220000in}{0.140000in}}%
\pgfpathlineto{\pgfqpoint{0.000000in}{0.140000in}}%
\pgfpathlineto{\pgfqpoint{0.000000in}{0.000000in}}%
\pgfpathclose%
\pgfusepath{fill}%
\end{pgfscope}%
\begin{pgfscope}%
\pgfpathrectangle{\pgfqpoint{0.000000in}{0.000000in}}{\pgfqpoint{0.220000in}{0.140000in}}%
\pgfusepath{clip}%
\pgfsetbuttcap%
\pgfsetroundjoin%
\definecolor{currentfill}{rgb}{0.246811,0.283237,0.535941}%
\pgfsetfillcolor{currentfill}%
\pgfsetlinewidth{0.351312pt}%
\definecolor{currentstroke}{rgb}{1.000000,1.000000,1.000000}%
\pgfsetstrokecolor{currentstroke}%
\pgfsetdash{}{0pt}%
\pgfsys@defobject{currentmarker}{\pgfqpoint{-0.036747in}{-0.036747in}}{\pgfqpoint{0.036747in}{0.036747in}}{%
\pgfpathmoveto{\pgfqpoint{-0.036747in}{-0.036747in}}%
\pgfpathlineto{\pgfqpoint{0.036747in}{-0.036747in}}%
\pgfpathlineto{\pgfqpoint{0.036747in}{0.036747in}}%
\pgfpathlineto{\pgfqpoint{-0.036747in}{0.036747in}}%
\pgfpathlineto{\pgfqpoint{-0.036747in}{-0.036747in}}%
\pgfpathclose%
\pgfusepath{stroke,fill}%
}%
\begin{pgfscope}%
\pgfsys@transformshift{0.110000in}{0.070000in}%
\pgfsys@useobject{currentmarker}{}%
\end{pgfscope}%
\end{pgfscope}%
\end{pgfpicture}%
\makeatother%
\endgroup%
} & $\mathrm{HM}$ & Hierarchical MLP mixer \\
\raisebox{-0.25em}{
\begingroup%
\makeatletter%
\begin{pgfpicture}%
\pgfpathrectangle{\pgfpointorigin}{\pgfqpoint{0.220000in}{0.140000in}}%
\pgfusepath{use as bounding box, clip}%
\begin{pgfscope}%
\pgfsetbuttcap%
\pgfsetmiterjoin%
\definecolor{currentfill}{rgb}{1.000000,1.000000,1.000000}%
\pgfsetfillcolor{currentfill}%
\pgfsetlinewidth{0.000000pt}%
\definecolor{currentstroke}{rgb}{1.000000,1.000000,1.000000}%
\pgfsetstrokecolor{currentstroke}%
\pgfsetdash{}{0pt}%
\pgfpathmoveto{\pgfqpoint{0.000000in}{0.000000in}}%
\pgfpathlineto{\pgfqpoint{0.220000in}{0.000000in}}%
\pgfpathlineto{\pgfqpoint{0.220000in}{0.140000in}}%
\pgfpathlineto{\pgfqpoint{0.000000in}{0.140000in}}%
\pgfpathlineto{\pgfqpoint{0.000000in}{0.000000in}}%
\pgfpathclose%
\pgfusepath{fill}%
\end{pgfscope}%
\begin{pgfscope}%
\pgfpathrectangle{\pgfqpoint{0.000000in}{0.000000in}}{\pgfqpoint{0.220000in}{0.140000in}}%
\pgfusepath{clip}%
\pgfsetbuttcap%
\pgfsetroundjoin%
\definecolor{currentfill}{rgb}{0.199430,0.387607,0.554642}%
\pgfsetfillcolor{currentfill}%
\pgfsetlinewidth{0.351312pt}%
\definecolor{currentstroke}{rgb}{1.000000,1.000000,1.000000}%
\pgfsetstrokecolor{currentstroke}%
\pgfsetdash{}{0pt}%
\pgfsys@defobject{currentmarker}{\pgfqpoint{-0.036747in}{-0.036747in}}{\pgfqpoint{0.036747in}{0.036747in}}{%
\pgfpathmoveto{\pgfqpoint{0.000000in}{-0.036747in}}%
\pgfpathcurveto{\pgfqpoint{0.009745in}{-0.036747in}}{\pgfqpoint{0.019093in}{-0.032875in}}{\pgfqpoint{0.025984in}{-0.025984in}}%
\pgfpathcurveto{\pgfqpoint{0.032875in}{-0.019093in}}{\pgfqpoint{0.036747in}{-0.009745in}}{\pgfqpoint{0.036747in}{0.000000in}}%
\pgfpathcurveto{\pgfqpoint{0.036747in}{0.009745in}}{\pgfqpoint{0.032875in}{0.019093in}}{\pgfqpoint{0.025984in}{0.025984in}}%
\pgfpathcurveto{\pgfqpoint{0.019093in}{0.032875in}}{\pgfqpoint{0.009745in}{0.036747in}}{\pgfqpoint{0.000000in}{0.036747in}}%
\pgfpathcurveto{\pgfqpoint{-0.009745in}{0.036747in}}{\pgfqpoint{-0.019093in}{0.032875in}}{\pgfqpoint{-0.025984in}{0.025984in}}%
\pgfpathcurveto{\pgfqpoint{-0.032875in}{0.019093in}}{\pgfqpoint{-0.036747in}{0.009745in}}{\pgfqpoint{-0.036747in}{0.000000in}}%
\pgfpathcurveto{\pgfqpoint{-0.036747in}{-0.009745in}}{\pgfqpoint{-0.032875in}{-0.019093in}}{\pgfqpoint{-0.025984in}{-0.025984in}}%
\pgfpathcurveto{\pgfqpoint{-0.019093in}{-0.032875in}}{\pgfqpoint{-0.009745in}{-0.036747in}}{\pgfqpoint{0.000000in}{-0.036747in}}%
\pgfpathlineto{\pgfqpoint{0.000000in}{-0.036747in}}%
\pgfpathclose%
\pgfusepath{stroke,fill}%
}%
\begin{pgfscope}%
\pgfsys@transformshift{0.110000in}{0.070000in}%
\pgfsys@useobject{currentmarker}{}%
\end{pgfscope}%
\end{pgfscope}%
\end{pgfpicture}%
\makeatother%
\endgroup%
} & $\mathrm{RNN}$ & Site-wise recurrent neural network \\
\raisebox{-0.25em}{
\begingroup%
\makeatletter%
\begin{pgfpicture}%
\pgfpathrectangle{\pgfpointorigin}{\pgfqpoint{0.220000in}{0.140000in}}%
\pgfusepath{use as bounding box, clip}%
\begin{pgfscope}%
\pgfsetbuttcap%
\pgfsetmiterjoin%
\definecolor{currentfill}{rgb}{1.000000,1.000000,1.000000}%
\pgfsetfillcolor{currentfill}%
\pgfsetlinewidth{0.000000pt}%
\definecolor{currentstroke}{rgb}{1.000000,1.000000,1.000000}%
\pgfsetstrokecolor{currentstroke}%
\pgfsetdash{}{0pt}%
\pgfpathmoveto{\pgfqpoint{0.000000in}{0.000000in}}%
\pgfpathlineto{\pgfqpoint{0.220000in}{0.000000in}}%
\pgfpathlineto{\pgfqpoint{0.220000in}{0.140000in}}%
\pgfpathlineto{\pgfqpoint{0.000000in}{0.140000in}}%
\pgfpathlineto{\pgfqpoint{0.000000in}{0.000000in}}%
\pgfpathclose%
\pgfusepath{fill}%
\end{pgfscope}%
\begin{pgfscope}%
\pgfpathrectangle{\pgfqpoint{0.000000in}{0.000000in}}{\pgfqpoint{0.220000in}{0.140000in}}%
\pgfusepath{clip}%
\pgfsetbuttcap%
\pgfsetroundjoin%
\definecolor{currentfill}{rgb}{0.159194,0.482237,0.558073}%
\pgfsetfillcolor{currentfill}%
\pgfsetlinewidth{0.351312pt}%
\definecolor{currentstroke}{rgb}{1.000000,1.000000,1.000000}%
\pgfsetstrokecolor{currentstroke}%
\pgfsetdash{}{0pt}%
\pgfsys@defobject{currentmarker}{\pgfqpoint{-0.036747in}{-0.036747in}}{\pgfqpoint{0.036747in}{0.036747in}}{%
\pgfpathmoveto{\pgfqpoint{0.000000in}{0.036747in}}%
\pgfpathlineto{\pgfqpoint{-0.036747in}{-0.036747in}}%
\pgfpathlineto{\pgfqpoint{0.036747in}{-0.036747in}}%
\pgfpathlineto{\pgfqpoint{0.000000in}{0.036747in}}%
\pgfpathclose%
\pgfusepath{stroke,fill}%
}%
\begin{pgfscope}%
\pgfsys@transformshift{0.110000in}{0.070000in}%
\pgfsys@useobject{currentmarker}{}%
\end{pgfscope}%
\end{pgfscope}%
\end{pgfpicture}%
\makeatother%
\endgroup%
} & $\mathrm{PRNN}$ & Patched recurrent neural network \\
\raisebox{-0.25em}{
\begingroup%
\makeatletter%
\begin{pgfpicture}%
\pgfpathrectangle{\pgfpointorigin}{\pgfqpoint{0.220000in}{0.140000in}}%
\pgfusepath{use as bounding box, clip}%
\begin{pgfscope}%
\pgfsetbuttcap%
\pgfsetmiterjoin%
\definecolor{currentfill}{rgb}{1.000000,1.000000,1.000000}%
\pgfsetfillcolor{currentfill}%
\pgfsetlinewidth{0.000000pt}%
\definecolor{currentstroke}{rgb}{1.000000,1.000000,1.000000}%
\pgfsetstrokecolor{currentstroke}%
\pgfsetdash{}{0pt}%
\pgfpathmoveto{\pgfqpoint{0.000000in}{0.000000in}}%
\pgfpathlineto{\pgfqpoint{0.220000in}{0.000000in}}%
\pgfpathlineto{\pgfqpoint{0.220000in}{0.140000in}}%
\pgfpathlineto{\pgfqpoint{0.000000in}{0.140000in}}%
\pgfpathlineto{\pgfqpoint{0.000000in}{0.000000in}}%
\pgfpathclose%
\pgfusepath{fill}%
\end{pgfscope}%
\begin{pgfscope}%
\pgfpathrectangle{\pgfqpoint{0.000000in}{0.000000in}}{\pgfqpoint{0.220000in}{0.140000in}}%
\pgfusepath{clip}%
\pgfsetbuttcap%
\pgfsetroundjoin%
\definecolor{currentfill}{rgb}{0.125394,0.574318,0.549086}%
\pgfsetfillcolor{currentfill}%
\pgfsetlinewidth{0.351312pt}%
\definecolor{currentstroke}{rgb}{1.000000,1.000000,1.000000}%
\pgfsetstrokecolor{currentstroke}%
\pgfsetdash{}{0pt}%
\pgfsys@defobject{currentmarker}{\pgfqpoint{-0.036747in}{-0.036747in}}{\pgfqpoint{0.036747in}{0.036747in}}{%
\pgfpathmoveto{\pgfqpoint{-0.036747in}{-0.036747in}}%
\pgfpathlineto{\pgfqpoint{0.036747in}{-0.036747in}}%
\pgfpathlineto{\pgfqpoint{0.036747in}{0.036747in}}%
\pgfpathlineto{\pgfqpoint{-0.036747in}{0.036747in}}%
\pgfpathlineto{\pgfqpoint{-0.036747in}{-0.036747in}}%
\pgfpathclose%
\pgfusepath{stroke,fill}%
}%
\begin{pgfscope}%
\pgfsys@transformshift{0.110000in}{0.070000in}%
\pgfsys@useobject{currentmarker}{}%
\end{pgfscope}%
\end{pgfscope}%
\end{pgfpicture}%
\makeatother%
\endgroup%
} & $\mathrm{HRNN}$ & Hierarchical recurrent neural network \\
\raisebox{-0.25em}{
\begingroup%
\makeatletter%
\begin{pgfpicture}%
\pgfpathrectangle{\pgfpointorigin}{\pgfqpoint{0.220000in}{0.140000in}}%
\pgfusepath{use as bounding box, clip}%
\begin{pgfscope}%
\pgfsetbuttcap%
\pgfsetmiterjoin%
\definecolor{currentfill}{rgb}{1.000000,1.000000,1.000000}%
\pgfsetfillcolor{currentfill}%
\pgfsetlinewidth{0.000000pt}%
\definecolor{currentstroke}{rgb}{1.000000,1.000000,1.000000}%
\pgfsetstrokecolor{currentstroke}%
\pgfsetdash{}{0pt}%
\pgfpathmoveto{\pgfqpoint{0.000000in}{0.000000in}}%
\pgfpathlineto{\pgfqpoint{0.220000in}{0.000000in}}%
\pgfpathlineto{\pgfqpoint{0.220000in}{0.140000in}}%
\pgfpathlineto{\pgfqpoint{0.000000in}{0.140000in}}%
\pgfpathlineto{\pgfqpoint{0.000000in}{0.000000in}}%
\pgfpathclose%
\pgfusepath{fill}%
\end{pgfscope}%
\begin{pgfscope}%
\pgfpathrectangle{\pgfqpoint{0.000000in}{0.000000in}}{\pgfqpoint{0.220000in}{0.140000in}}%
\pgfusepath{clip}%
\pgfsetbuttcap%
\pgfsetroundjoin%
\definecolor{currentfill}{rgb}{0.140210,0.665859,0.513427}%
\pgfsetfillcolor{currentfill}%
\pgfsetlinewidth{0.351312pt}%
\definecolor{currentstroke}{rgb}{1.000000,1.000000,1.000000}%
\pgfsetstrokecolor{currentstroke}%
\pgfsetdash{}{0pt}%
\pgfsys@defobject{currentmarker}{\pgfqpoint{-0.036747in}{-0.036747in}}{\pgfqpoint{0.036747in}{0.036747in}}{%
\pgfpathmoveto{\pgfqpoint{0.000000in}{-0.036747in}}%
\pgfpathcurveto{\pgfqpoint{0.009745in}{-0.036747in}}{\pgfqpoint{0.019093in}{-0.032875in}}{\pgfqpoint{0.025984in}{-0.025984in}}%
\pgfpathcurveto{\pgfqpoint{0.032875in}{-0.019093in}}{\pgfqpoint{0.036747in}{-0.009745in}}{\pgfqpoint{0.036747in}{0.000000in}}%
\pgfpathcurveto{\pgfqpoint{0.036747in}{0.009745in}}{\pgfqpoint{0.032875in}{0.019093in}}{\pgfqpoint{0.025984in}{0.025984in}}%
\pgfpathcurveto{\pgfqpoint{0.019093in}{0.032875in}}{\pgfqpoint{0.009745in}{0.036747in}}{\pgfqpoint{0.000000in}{0.036747in}}%
\pgfpathcurveto{\pgfqpoint{-0.009745in}{0.036747in}}{\pgfqpoint{-0.019093in}{0.032875in}}{\pgfqpoint{-0.025984in}{0.025984in}}%
\pgfpathcurveto{\pgfqpoint{-0.032875in}{0.019093in}}{\pgfqpoint{-0.036747in}{0.009745in}}{\pgfqpoint{-0.036747in}{0.000000in}}%
\pgfpathcurveto{\pgfqpoint{-0.036747in}{-0.009745in}}{\pgfqpoint{-0.032875in}{-0.019093in}}{\pgfqpoint{-0.025984in}{-0.025984in}}%
\pgfpathcurveto{\pgfqpoint{-0.019093in}{-0.032875in}}{\pgfqpoint{-0.009745in}{-0.036747in}}{\pgfqpoint{0.000000in}{-0.036747in}}%
\pgfpathlineto{\pgfqpoint{0.000000in}{-0.036747in}}%
\pgfpathclose%
\pgfusepath{stroke,fill}%
}%
\begin{pgfscope}%
\pgfsys@transformshift{0.110000in}{0.070000in}%
\pgfsys@useobject{currentmarker}{}%
\end{pgfscope}%
\end{pgfscope}%
\end{pgfpicture}%
\makeatother%
\endgroup%
} & $\mathrm{TF}$ & Site-wise transformer \\
\raisebox{-0.25em}{
\begingroup%
\makeatletter%
\begin{pgfpicture}%
\pgfpathrectangle{\pgfpointorigin}{\pgfqpoint{0.220000in}{0.140000in}}%
\pgfusepath{use as bounding box, clip}%
\begin{pgfscope}%
\pgfsetbuttcap%
\pgfsetmiterjoin%
\definecolor{currentfill}{rgb}{1.000000,1.000000,1.000000}%
\pgfsetfillcolor{currentfill}%
\pgfsetlinewidth{0.000000pt}%
\definecolor{currentstroke}{rgb}{1.000000,1.000000,1.000000}%
\pgfsetstrokecolor{currentstroke}%
\pgfsetdash{}{0pt}%
\pgfpathmoveto{\pgfqpoint{0.000000in}{0.000000in}}%
\pgfpathlineto{\pgfqpoint{0.220000in}{0.000000in}}%
\pgfpathlineto{\pgfqpoint{0.220000in}{0.140000in}}%
\pgfpathlineto{\pgfqpoint{0.000000in}{0.140000in}}%
\pgfpathlineto{\pgfqpoint{0.000000in}{0.000000in}}%
\pgfpathclose%
\pgfusepath{fill}%
\end{pgfscope}%
\begin{pgfscope}%
\pgfpathrectangle{\pgfqpoint{0.000000in}{0.000000in}}{\pgfqpoint{0.220000in}{0.140000in}}%
\pgfusepath{clip}%
\pgfsetbuttcap%
\pgfsetroundjoin%
\definecolor{currentfill}{rgb}{0.274149,0.751988,0.436601}%
\pgfsetfillcolor{currentfill}%
\pgfsetlinewidth{0.351312pt}%
\definecolor{currentstroke}{rgb}{1.000000,1.000000,1.000000}%
\pgfsetstrokecolor{currentstroke}%
\pgfsetdash{}{0pt}%
\pgfsys@defobject{currentmarker}{\pgfqpoint{-0.036747in}{-0.036747in}}{\pgfqpoint{0.036747in}{0.036747in}}{%
\pgfpathmoveto{\pgfqpoint{0.000000in}{0.036747in}}%
\pgfpathlineto{\pgfqpoint{-0.036747in}{-0.036747in}}%
\pgfpathlineto{\pgfqpoint{0.036747in}{-0.036747in}}%
\pgfpathlineto{\pgfqpoint{0.000000in}{0.036747in}}%
\pgfpathclose%
\pgfusepath{stroke,fill}%
}%
\begin{pgfscope}%
\pgfsys@transformshift{0.110000in}{0.070000in}%
\pgfsys@useobject{currentmarker}{}%
\end{pgfscope}%
\end{pgfscope}%
\end{pgfpicture}%
\makeatother%
\endgroup%
} & $\mathrm{PTF}$ & Patched transformer \\
\raisebox{-0.25em}{
\begingroup%
\makeatletter%
\begin{pgfpicture}%
\pgfpathrectangle{\pgfpointorigin}{\pgfqpoint{0.220000in}{0.140000in}}%
\pgfusepath{use as bounding box, clip}%
\begin{pgfscope}%
\pgfsetbuttcap%
\pgfsetmiterjoin%
\definecolor{currentfill}{rgb}{1.000000,1.000000,1.000000}%
\pgfsetfillcolor{currentfill}%
\pgfsetlinewidth{0.000000pt}%
\definecolor{currentstroke}{rgb}{1.000000,1.000000,1.000000}%
\pgfsetstrokecolor{currentstroke}%
\pgfsetdash{}{0pt}%
\pgfpathmoveto{\pgfqpoint{0.000000in}{0.000000in}}%
\pgfpathlineto{\pgfqpoint{0.220000in}{0.000000in}}%
\pgfpathlineto{\pgfqpoint{0.220000in}{0.140000in}}%
\pgfpathlineto{\pgfqpoint{0.000000in}{0.140000in}}%
\pgfpathlineto{\pgfqpoint{0.000000in}{0.000000in}}%
\pgfpathclose%
\pgfusepath{fill}%
\end{pgfscope}%
\begin{pgfscope}%
\pgfpathrectangle{\pgfqpoint{0.000000in}{0.000000in}}{\pgfqpoint{0.220000in}{0.140000in}}%
\pgfusepath{clip}%
\pgfsetbuttcap%
\pgfsetroundjoin%
\definecolor{currentfill}{rgb}{0.487026,0.823929,0.312321}%
\pgfsetfillcolor{currentfill}%
\pgfsetlinewidth{0.351312pt}%
\definecolor{currentstroke}{rgb}{1.000000,1.000000,1.000000}%
\pgfsetstrokecolor{currentstroke}%
\pgfsetdash{}{0pt}%
\pgfsys@defobject{currentmarker}{\pgfqpoint{-0.051967in}{-0.051967in}}{\pgfqpoint{0.051967in}{0.051967in}}{%
\pgfpathmoveto{\pgfqpoint{-0.000000in}{-0.051967in}}%
\pgfpathlineto{\pgfqpoint{0.051967in}{0.000000in}}%
\pgfpathlineto{\pgfqpoint{0.000000in}{0.051967in}}%
\pgfpathlineto{\pgfqpoint{-0.051967in}{0.000000in}}%
\pgfpathlineto{\pgfqpoint{-0.000000in}{-0.051967in}}%
\pgfpathclose%
\pgfusepath{stroke,fill}%
}%
\begin{pgfscope}%
\pgfsys@transformshift{0.110000in}{0.070000in}%
\pgfsys@useobject{currentmarker}{}%
\end{pgfscope}%
\end{pgfscope}%
\end{pgfpicture}%
\makeatother%
\endgroup%
} & $\mathrm{HTF{+}M}$ & Hierarchical transformer with MLP mixer model subsampler \\
\raisebox{-0.25em}{
\begingroup%
\makeatletter%
\begin{pgfpicture}%
\pgfpathrectangle{\pgfpointorigin}{\pgfqpoint{0.220000in}{0.140000in}}%
\pgfusepath{use as bounding box, clip}%
\begin{pgfscope}%
\pgfsetbuttcap%
\pgfsetmiterjoin%
\definecolor{currentfill}{rgb}{1.000000,1.000000,1.000000}%
\pgfsetfillcolor{currentfill}%
\pgfsetlinewidth{0.000000pt}%
\definecolor{currentstroke}{rgb}{1.000000,1.000000,1.000000}%
\pgfsetstrokecolor{currentstroke}%
\pgfsetdash{}{0pt}%
\pgfpathmoveto{\pgfqpoint{0.000000in}{0.000000in}}%
\pgfpathlineto{\pgfqpoint{0.220000in}{0.000000in}}%
\pgfpathlineto{\pgfqpoint{0.220000in}{0.140000in}}%
\pgfpathlineto{\pgfqpoint{0.000000in}{0.140000in}}%
\pgfpathlineto{\pgfqpoint{0.000000in}{0.000000in}}%
\pgfpathclose%
\pgfusepath{fill}%
\end{pgfscope}%
\begin{pgfscope}%
\pgfpathrectangle{\pgfqpoint{0.000000in}{0.000000in}}{\pgfqpoint{0.220000in}{0.140000in}}%
\pgfusepath{clip}%
\pgfsetbuttcap%
\pgfsetroundjoin%
\definecolor{currentfill}{rgb}{0.741388,0.873449,0.149561}%
\pgfsetfillcolor{currentfill}%
\pgfsetlinewidth{0.351312pt}%
\definecolor{currentstroke}{rgb}{1.000000,1.000000,1.000000}%
\pgfsetstrokecolor{currentstroke}%
\pgfsetdash{}{0pt}%
\pgfsys@defobject{currentmarker}{\pgfqpoint{-0.036747in}{-0.036747in}}{\pgfqpoint{0.036747in}{0.036747in}}{%
\pgfpathmoveto{\pgfqpoint{-0.012249in}{-0.036747in}}%
\pgfpathlineto{\pgfqpoint{0.012249in}{-0.036747in}}%
\pgfpathlineto{\pgfqpoint{0.012249in}{-0.012249in}}%
\pgfpathlineto{\pgfqpoint{0.036747in}{-0.012249in}}%
\pgfpathlineto{\pgfqpoint{0.036747in}{0.012249in}}%
\pgfpathlineto{\pgfqpoint{0.012249in}{0.012249in}}%
\pgfpathlineto{\pgfqpoint{0.012249in}{0.036747in}}%
\pgfpathlineto{\pgfqpoint{-0.012249in}{0.036747in}}%
\pgfpathlineto{\pgfqpoint{-0.012249in}{0.012249in}}%
\pgfpathlineto{\pgfqpoint{-0.036747in}{0.012249in}}%
\pgfpathlineto{\pgfqpoint{-0.036747in}{-0.012249in}}%
\pgfpathlineto{\pgfqpoint{-0.012249in}{-0.012249in}}%
\pgfpathlineto{\pgfqpoint{-0.012249in}{-0.036747in}}%
\pgfpathclose%
\pgfusepath{stroke,fill}%
}%
\begin{pgfscope}%
\pgfsys@transformshift{0.110000in}{0.070000in}%
\pgfsys@useobject{currentmarker}{}%
\end{pgfscope}%
\end{pgfscope}%
\end{pgfpicture}%
\makeatother%
\endgroup%
} & $\mathrm{HTF{+}R}$ & Hierarchical transformer with RNN subsampler \\
\end{tabular}

  \end{ruledtabular}
\end{table*}

\begin{figure*}
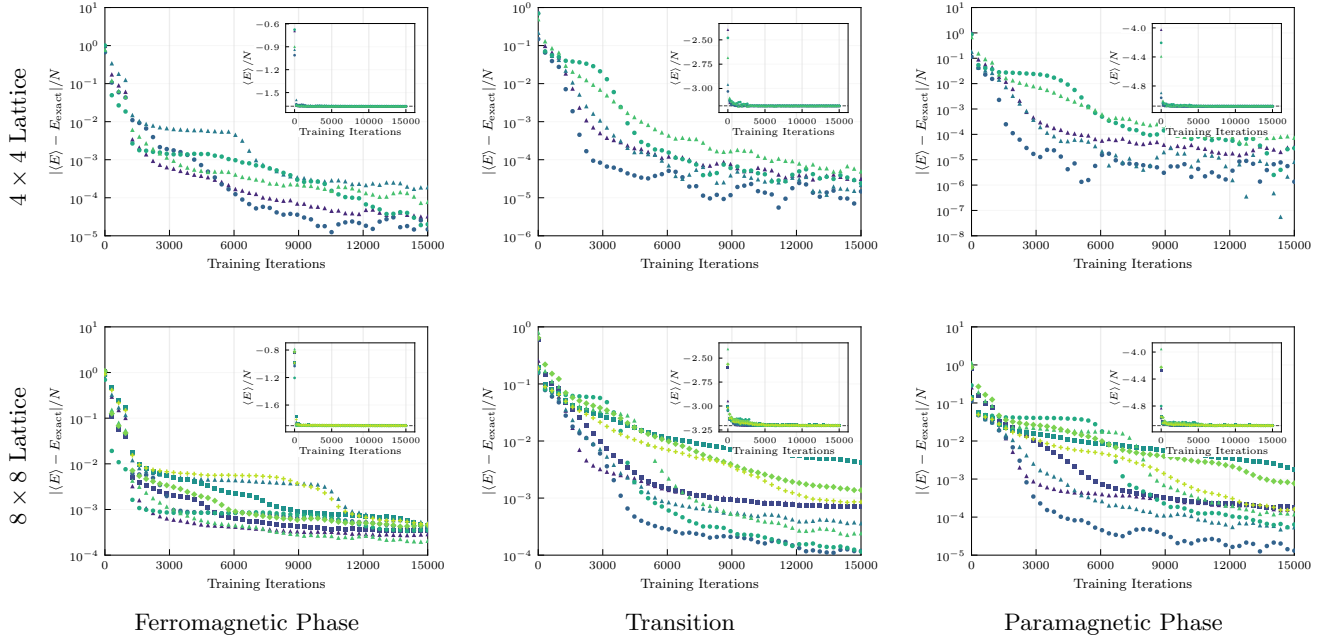

\centering
\begin{tikzpicture}
\begin{groupplot}[
    group style={
        group size=3 by 2,
        horizontal sep=5pt,
        vertical sep=8pt,
        x descriptions at=edge bottom,
        y descriptions at=edge left
    },
    scale only axis,
    width=0.31\textwidth,
    height=0.22\textwidth,
    xmin=0, xmax=1,
    ymin=0, ymax=1,
    xtick=\empty,
    ytick=\empty,
    enlargelimits=false,
    axis line style={draw=none},
    tick style={draw=none},
]
\nextgroupplot[ylabel={$4 \times 4$ Lattice}]
\node[inner sep=0pt] at (rel axis cs:0.5,0.5)
  {\resizebox{0.31\textwidth}{!}{\input{tfim_4x4_h_1.pgf}}};

\nextgroupplot
\node[inner sep=0pt] at (rel axis cs:0.5,0.5)
  {\resizebox{0.31\textwidth}{!}{\input{tfim_4x4_h_3_044.pgf}}};

\nextgroupplot
\node[inner sep=0pt] at (rel axis cs:0.5,0.5)
  {\resizebox{0.31\textwidth}{!}{\input{tfim_4x4_h_5.pgf}}};

\nextgroupplot[ylabel={$8 \times 8$ Lattice}, xlabel={Ferromagnetic Phase}]
\node[inner sep=0pt] at (rel axis cs:0.5,0.5)
  {\resizebox{0.31\textwidth}{!}{\input{tfim_8x8_h_1.pgf}}};

\nextgroupplot[xlabel={Transition}]
\node[inner sep=0pt] at (rel axis cs:0.5,0.5)
  {\resizebox{0.31\textwidth}{!}{\input{tfim_8x8_h_3_044.pgf}}};

\nextgroupplot[xlabel={Paramagnetic Phase}]
\node[inner sep=0pt] at (rel axis cs:0.5,0.5)
  {\resizebox{0.31\textwidth}{!}{\input{tfim_8x8_h_5.pgf}}};

\end{groupplot}
\end{tikzpicture}
\caption{Search for TFIM ground states in square lattices of $4\times 4$ and $8\times 8$ sites at $h = 1$, $h=3.044$, and $h=5$. The nine different ANN architectures (different colors) are variationally trained to represent the ground state. The main plot shows the evolution of the absolute energy-density error $\left|\left\langle E\right\rangle - E_{\mathrm{exact}}\right|/N$ for all ANN architectures during training for $15{,}000$ iterations where we compare to the exact ground state energy density obtained via tensor network simulations~\cite{itensor}. The inset shows the corresponding energy density $\left\langle E\right\rangle /N$, where all data points start to overlap quickly with the exact energy as indicated by a black horizontal line. Data points correspond to averages of the preceding $1000$ training iterations. Hierarchical ANN architectures use initial input patches of $4\times 4$ sites. The legend for this figure is provided in \Tab{architecture-plot-key}.}
\label{fig:all_tfim_plots}
\end{figure*}

\begin{figure*}[t]
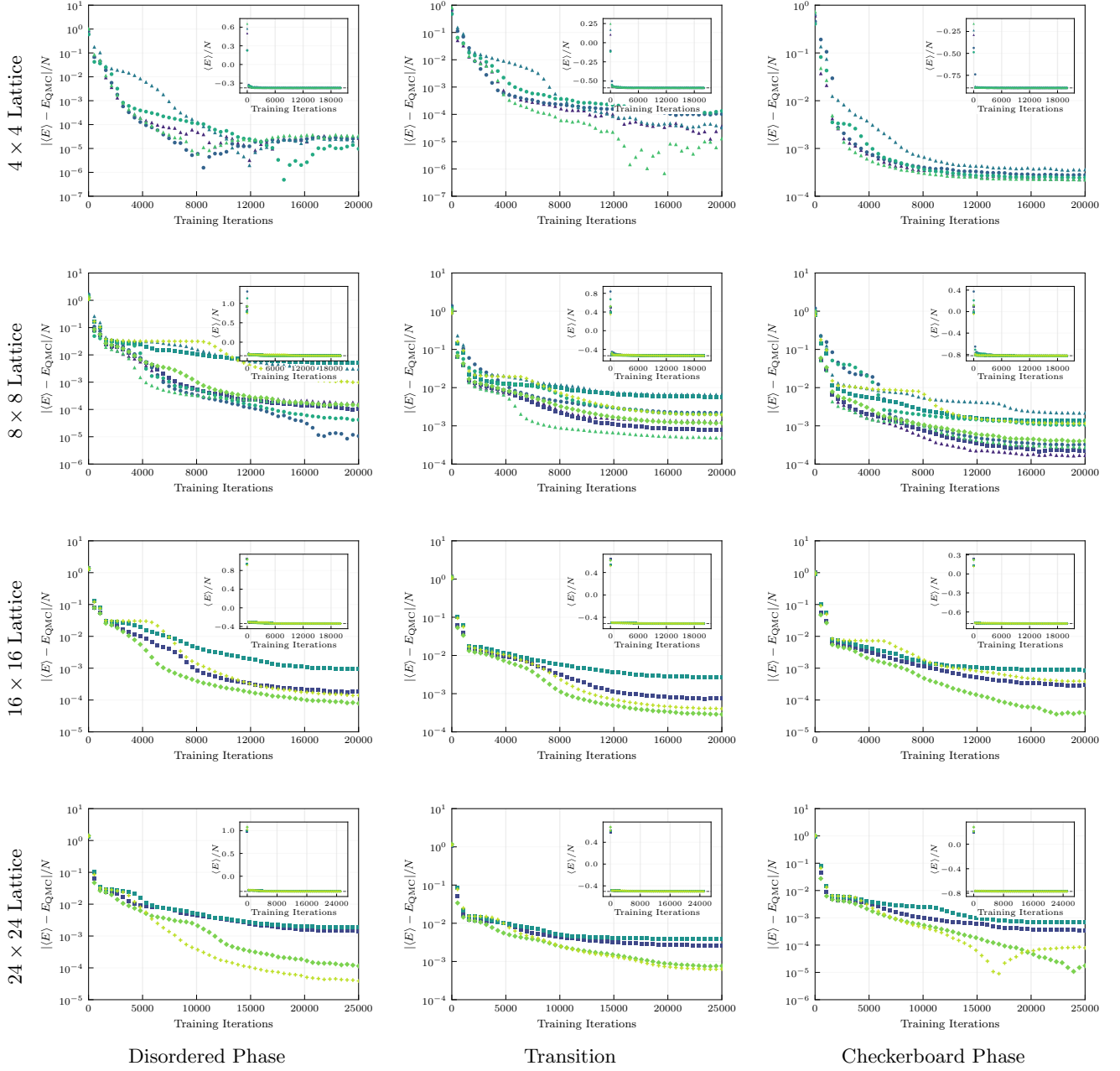

\centering
\begin{tikzpicture}
\begin{groupplot}[
    group style={
        group size=3 by 4,
        horizontal sep=5pt,
        vertical sep=8pt,
        x descriptions at=edge bottom,
        y descriptions at=edge left
    },
    scale only axis,
    width=0.31\textwidth,
    height=0.22\textwidth,
    xmin=0, xmax=1,
    ymin=0, ymax=1,
    xtick=\empty,
    ytick=\empty,
    enlargelimits=false,
    axis line style={draw=none},
    tick style={draw=none},
]
\nextgroupplot[ylabel={$4 \times 4$ Lattice}]
\node[inner sep=0pt] at (rel axis cs:0.5,0.5)
  {\resizebox{0.31\textwidth}{!}{\input{ryd_4x4_delta_0_4.pgf}}};

\nextgroupplot
\node[inner sep=0pt] at (rel axis cs:0.5,0.5)
  {\resizebox{0.31\textwidth}{!}{\input{ryd_4x4_delta_1_2.pgf}}};

\nextgroupplot
\node[inner sep=0pt] at (rel axis cs:0.5,0.5)
  {\resizebox{0.31\textwidth}{!}{\input{ryd_4x4_delta_2_0.pgf}}};

\nextgroupplot[ylabel={$8 \times 8$ Lattice}]
\node[inner sep=0pt] at (rel axis cs:0.5,0.5)
  {\resizebox{0.31\textwidth}{!}{\input{ryd_8x8_delta_0_4.pgf}}};

\nextgroupplot
\node[inner sep=0pt] at (rel axis cs:0.5,0.5)
  {\resizebox{0.31\textwidth}{!}{\input{ryd_8x8_delta_1_2.pgf}}};
  
\nextgroupplot
\node[inner sep=0pt] at (rel axis cs:0.5,0.5)
  {\resizebox{0.31\textwidth}{!}{\input{ryd_8x8_delta_2_0.pgf}}};

\nextgroupplot[ylabel={$16 \times 16$ Lattice}]
\node[inner sep=0pt] at (rel axis cs:0.5,0.5)
  {\resizebox{0.31\textwidth}{!}{\input{ryd_16x16_delta_0_4.pgf}}};

\nextgroupplot
\node[inner sep=0pt] at (rel axis cs:0.5,0.5)
  {\resizebox{0.31\textwidth}{!}{\input{ryd_16x16_delta_1_2.pgf}}};

\nextgroupplot
\node[inner sep=0pt] at (rel axis cs:0.5,0.5)
  {\resizebox{0.31\textwidth}{!}{\input{ryd_16x16_delta_2_0.pgf}}};

\nextgroupplot[ylabel={$24 \times 24$ Lattice}, xlabel={Disordered Phase}]
\node[inner sep=0pt] at (rel axis cs:0.5,0.5)
  {\resizebox{0.31\textwidth}{!}{\input{ryd_24x24_delta_0_4.pgf}}};

\nextgroupplot[xlabel={Transition}]
\node[inner sep=0pt] at (rel axis cs:0.5,0.5)
  {\resizebox{0.31\textwidth}{!}{\input{ryd_24x24_delta_1_2.pgf}}};

\nextgroupplot[xlabel={Checkerboard Phase}]
\node[inner sep=0pt] at (rel axis cs:0.5,0.5)
  {\resizebox{0.31\textwidth}{!}{\input{ryd_24x24_delta_2_0.pgf}}};
\end{groupplot}
\end{tikzpicture}
\caption{Search for ground states in two-dimensional Rydberg atom arrays arranged on square lattices of $4\times 4$, $8\times 8$, $16\times 16$, and $24\times 24$ atoms at $\delta = 0.4$, $\delta=1.2$, and $\delta=2$. The nine different ANN architectures (different colors) are variationally trained to represent the ground state. The main plot shows the evolution of the absolute energy-density error $\left|\left\langle E\right\rangle - E_{\mathrm{QMC}}\right|/N$ for all ANN architectures during training for $20{,}000$ iterations for $N=4\times 4$, $N=8\times 8$, and $N=16\times 16$ and $25{,}000$ iterations for $N=24\times 24$, where we compare to the exact ground state energy density estimated via quantum Monte Carlo simulations~\cite{merali_stochastic_2021}. The inset shows the corresponding energy density $\left\langle E\right\rangle /N$, where all data points start to overlap quickly with the exact energy indicated by a black horizontal line. Data points correspond to averages of the preceding $1000$ training iterations. Hierarchical ANN architectures use initial input patches of $4\times 4$ sites. The legend for this figure is provided in \Tab{architecture-plot-key}.}
\label{fig:all_rydberg_plots}
\end{figure*}

In \Fig{4}, \Fig{5}, and \Fig{6}, we evaluate the energy expectation values obtained after training all considered ANN architectures for $15{,}000$ (TFIM), $20{,}000$ (Rydberg array with $N=4\times 4$, $N=8\times 8$, and $N=16\times 16$), or $25{,}000$ (Rydberg array with $N=24\times 24$) iterations while considering different points in the phase diagram.
Here, we provide the corresponding training curves, where we plot the energy density error as a function of training iterations for the two-dimensional TFIM in \Fig{all_tfim_plots} and for the two-dimensional Rydberg atom array in \Fig{all_rydberg_plots}.
For each we show all settings considered in the manuscript, i.e., systems of $N=4\times 4$ and $N=8\times 8$ particles in both figures and additionally systems of $N=16\times 16$ and $N=24\times 24$ atoms in \Fig{all_rydberg_plots}.
We further show the three considered points in the phase diagram, which is $h=1$, $h=3.044$, and $h=5$ for the TFIM and $\delta=0.4$, $\delta=1.2$, and $\delta=2$ in the Rydberg atom array.
In both figures, main plots show the energy density error, while insets show the corresponding energy densities obtained from the ANN simulations.
Exact energies are obtained via tensor network simulations for the TFIM~\cite{itensor} and quantum Monte Carlo simulations for Rydberg atom arrays~\cite{merali_stochastic_2021}.
The nine different network architectures, together with their nomenclature and their visual representation in all figures throughout the manuscript, are summarized in \Tab{architecture-plot-key}.
All panels show that for any configuration, all network architectures converge smoothly to a final energy density value within the evaluated number of training iterations, justifying our considerations in the main text.

\section{Computational Cost (FLOPs/MACs)}
\label{app:computational-cost}

\begin{table*}[htbp]
  \caption{Trainable parameter counts and computational cost per forward pass (FLOPs) for all considered ANN architectures applied to find ground states of different system sizes in the TFIM and Rydberg Hamiltonian.}
  \label{tab:full-flops}
  \begin{ruledtabular}
  \begin{tabular}{lllrr}
    Hamiltonian & System size & Architecture & Trainable parameters
      & FLOPs per forward pass \\
    \hline
            \multirow{14}{*}{TFIM} & \multirow{5}{*}{$4\times4$} & $\mathrm{M}$ & $5{,}000$ & $48{,}000$ \\
     &  & $\mathrm{RNN}$ & $5{,}506$ & $156{,}672$ \\
     &  & $\mathrm{PRNN}$ & $4{,}996$ & $34{,}816$ \\
     &  & $\mathrm{TF}$ & $5{,}426$ & $197{,}632$ \\
     &  & $\mathrm{PTF}$ & $4{,}904$ & $35{,}840$ \\
    \cline{2-5}
     & \multirow{9}{*}{$8\times8$} & $\mathrm{M}$ & $10{,}080$ & $160{,}768$\\
     &  & $\mathrm{HM}$ & $10{,}080$ & $177{,}280$ \\
     &  & $\mathrm{RNN}$ & $9{,}914$ & $1{,}085{,}440$ \\
     &  & $\mathrm{PRNN}$ & $10{,}128$ & $294{,}912$ \\
     &  & $\mathrm{HRNN}$ & $9{,}936$ & $157{,}696$ \\
     &  & $\mathrm{TF}$ & $10{,}946$ & $1{,}970{,}176$ \\
     &  & $\mathrm{PTF}$ & $9{,}848$ & $327{,}680$ \\
     &  & $\mathrm{HTF{+}M}$ & $9{,}900$ & $114{,}688$ \\
     &  & $\mathrm{HTF{+}R}$ & $9{,}976$ & $147{,}456$ \\
    \hline
    \multirow{22}{*}{Rydberg} & \multirow{5}{*}{$4\times4$} & $\mathrm{M}$ & $5{,}000$ & $48{,}000$\\
     &  & $\mathrm{RNN}$ & $4{,}970$ & $145{,}920$ \\
     &  & $\mathrm{PRNN}$ & $4{,}996$ & $34{,}816$ \\
     &  & $\mathrm{TF}$ & $4{,}786$ & $164{,}864$ \\
     &  & $\mathrm{PTF}$ & $5{,}056$ & $36{,}864$ \\
    \cline{2-5}
     & \multirow{9}{*}{$8\times8$} & $\mathrm{M}$ & $10{,}080$ & $160{,}768$ \\
     &  & $\mathrm{HM}$ & $10{,}080$ & $177{,}280$ \\
     &  & $\mathrm{RNN}$ & $9{,}914$ & $1{,}085{,}440$ \\
     &  & $\mathrm{PRNN}$ & $9{,}960$ & $296{,}960$ \\
     &  & $\mathrm{HRNN}$ & $9{,}936$ & $120{,}832$ \\
     &  & $\mathrm{TF}$ & $9{,}778$ & $1{,}576{,}960$ \\
     &  & $\mathrm{PTF}$ & $9{,}960$ & $311{,}296$ \\
     &  & $\mathrm{HTF{+}M}$ & $10{,}168$ & $227{,}328$ \\
     &  & $\mathrm{HTF{+}R}$ & $10{,}256$ & $186{,}368$ \\
    \cline{2-5}
     & \multirow{4}{*}{$16\times16$} & $\mathrm{HM}$ & $19{,}896$ & $1{,}128{,}960$ \\
     &  & $\mathrm{HRNN}$ & $19{,}472$ & $1{,}400{,}832$ \\
     &  & $\mathrm{HTF{+}M}$ & $20{,}152$ & $1{,}458{,}176$ \\
     &  & $\mathrm{HTF{+}R}$ & $19{,}072$ & $1{,}431{,}552$ \\
    \cline{2-5}
     & \multirow{4}{*}{$24\times24$} & $\mathrm{HM}$ & $29{,}858$ & $2{,}312{,}192$\\
     &  & $\mathrm{HRNN}$ & $28{,}328$ & $3{,}646{,}464$ \\
     &  & $\mathrm{HTF{+}M}$ & $29{,}250$ & $3{,}264{,}512$ \\
     &  & $\mathrm{HTF{+}R}$ & $32{,}080$ & $4{,}050{,}944$ \\
  \end{tabular}
  \end{ruledtabular}
\end{table*}

\begin{table*}[htbp]
  \caption{Architecture configurations producing the parameter counts in
  Table~\ref{tab:full-flops}. $D$ is the primary (large patch architecture in the case of hierarchical models)
  layer count and $D_s$ the local sub-sampler layer count. 
  $C$ is the channel-mixing or embedding width, equal to the RNN hidden-state width. 
  $H$ is the token-mixing MLP width for MLP mixer models, input-projection width for RNNs, and feed-forward width for transformer models. 
  $p$ is the generated patch side length (the primary patch for hierarchical models), and $b$ the hierarchical sub-patch side length, both in terms of lattice sites. 
  Hierarchical primary and local networks use the same $C$ and $H$ dimensions with independent weights. 
  All recurrent cells are GRUs and all transformer models use four attention heads. 
    Dashes indicate non-applicable settings.}
  \label{tab:network-configurations}
  \begin{ruledtabular}
  \begin{tabular}{lllrrrrrr}
    Hamiltonian & System size & Architecture & $D$ & $C$ & $H$
      & $p$ & $b$ & $D_s$ \\
    \hline
    \multirow{14}{*}{TFIM}
      & \multirow{5}{*}{$4\times4$}
      & $\mathrm{M}$ & 10 & 12 & 12 & 2 & -- & -- \\
     & & $\mathrm{RNN}$  & 3 & 16 & 16 & 1 & -- & -- \\
     & & $\mathrm{PRNN}$ & 3 & 16 & 4  & 2 & -- & -- \\
     & & $\mathrm{TF}$   & 3 & 16 & 16 & 1 & -- & -- \\
     & & $\mathrm{PTF}$  & 3 & 16 & 8  & 2 & -- & -- \\
    \cline{2-9}
     & \multirow{9}{*}{$8\times8$}
       & $\mathrm{M}$       & 2 & 8  & 144 & 2 & -- & -- \\
     & & $\mathrm{HM}$      & 6 & 20 & 16  & 2 & 4 & 2 \\
     & & $\mathrm{RNN}$     & 3 & 16 & 72  & 1 & -- & -- \\
     & & $\mathrm{PRNN}$    & 3 & 16 & 80  & 2 & -- & -- \\
     & & $\mathrm{HRNN}$    & 3 & 16 & 16  & 2 & 4 & 2 \\
     & & $\mathrm{TF}$      & 3 & 16 & 64  & 1 & -- & -- \\
     & & $\mathrm{PTF}$     & 3 & 16 & 56  & 2 & -- & -- \\
     & & $\mathrm{HTF{+}M}$ & 4 & 16 & 24  & 2 & 4 & 1 \\
     & & $\mathrm{HTF{+}R}$ & 4 & 16 & 8   & 2 & 4 & 2 \\
    \hline
    \multirow{22}{*}{Rydberg}
      & \multirow{5}{*}{$4\times4$}
      & $\mathrm{M}$ & 10 & 12 & 12 & 2 & -- & -- \\
     & & $\mathrm{RNN}$  & 3 & 8  & 112 & 1 & -- & -- \\
     & & $\mathrm{PRNN}$ & 3 & 16 & 4   & 2 & -- & -- \\
     & & $\mathrm{TF}$   & 2 & 16 & 32  & 1 & -- & -- \\
     & & $\mathrm{PTF}$  & 2 & 16 & 32  & 2 & -- & -- \\
    \cline{2-9}
     & \multirow{9}{*}{$8\times8$}
       & $\mathrm{M}$       & 2 & 8  & 144 & 2 & -- & -- \\
     & & $\mathrm{HM}$      & 6 & 20 & 16  & 2 & 4 & 2 \\
     & & $\mathrm{RNN}$     & 3 & 16 & 72  & 1 & -- & -- \\
     & & $\mathrm{PRNN}$    & 3 & 8  & 256 & 2 & -- & -- \\
     & & $\mathrm{HRNN}$    & 4 & 16 & 16  & 2 & 4 & 1 \\
     & & $\mathrm{TF}$      & 2 & 16 & 96  & 1 & -- & -- \\
     & & $\mathrm{PTF}$     & 3 & 8  & 168 & 2 & -- & -- \\
     & & $\mathrm{HTF{+}M}$ & 2 & 16 & 64  & 2 & 4 & 2 \\
     & & $\mathrm{HTF{+}R}$ & 2 & 16 & 32  & 2 & 4 & 2 \\
    \cline{2-9}
     & \multirow{4}{*}{$16\times16$}
       & $\mathrm{HM}$      & 8 & 20 & 32 & 2 & 4 & 2 \\
     & & $\mathrm{HRNN}$    & 2 & 24 & 32 & 2 & 4 & 2 \\
     & & $\mathrm{HTF{+}M}$ & 2 & 32 & 24 & 2 & 4 & 2 \\
     & & $\mathrm{HTF{+}R}$ & 2 & 24 & 32 & 2 & 4 & 2 \\
    \cline{2-9}
     & \multirow{4}{*}{$24\times24$}
       & $\mathrm{HM}$      & 9 & 16 & 64 & 2 & 6 & 2 \\
     & & $\mathrm{HRNN}$    & 3 & 24 & 56 & 2 & 6 & 2 \\
     & & $\mathrm{HTF{+}M}$ & 3 & 32 & 32 & 2 & 6 & 2 \\
     & & $\mathrm{HTF{+}R}$ & 3 & 32 & 16 & 2 & 6 & 2 \\
  \end{tabular}
  \end{ruledtabular}
\end{table*}

In this work, we define a floating-point operation (FLOP) as a single scalar floating-point addition or multiplication. 
All FLOP counts follow standard dense linear-algebra conventions, i.e., the cost of matrix-matrix and matrix-vector products is computed analytically from tensor dimensions and operation structure. 
Nonlinear activation functions are treated as constant-cost operations per element and are included as a fixed number of scalar FLOPs per activation.

The numbers appearing in this manuscript include the FLOPs of a single forward pass in the considered ANN architecture, including linear projections, mixing layers, attention mechanisms, and embedding layers, as well as any architectural overhead introduced by patching or autoregressive sampling.
We do not include FLOPs used for gradient calculation via back-propagation, optimizer update, sample generation, observable evaluation, or data movement and memory access.

The forward-pass FLOPs provide a model-intrinsic, hardware-agnostic measure of computational complexity, enabling a fair comparison across the considered architectures.
Since sample generation and Hamiltonian evaluation depend on the studied physical model rather than the ANN architecture, we do not include these in our considerations.
We obtain the presented FLOP estimates using PyTorch's built-in context manager \texttt{FlopCounterMode} for tracking floating-point operations~\cite{cite_pytorch}.

A summary of the different ANN architectures is provided in \Tab{full-flops}, where we provide the trainable parameters and FLOPs per forward pass for all considered system sizes in the TFIM and Rydberg Hamiltonian.
In \Tab{network-configurations}, we further provide the detailed network setups to achieve the corresponding number of trainable parameters.


\clearpage

\begin{thebibliography}{68}%
\makeatletter
\providecommand \@ifxundefined [1]{%
 \@ifx{#1\undefined}
}%
\providecommand \@ifnum [1]{%
 \ifnum #1\expandafter \@firstoftwo
 \else \expandafter \@secondoftwo
 \fi
}%
\providecommand \@ifx [1]{%
 \ifx #1\expandafter \@firstoftwo
 \else \expandafter \@secondoftwo
 \fi
}%
\providecommand \natexlab [1]{#1}%
\providecommand \enquote  [1]{``#1''}%
\providecommand \bibnamefont  [1]{#1}%
\providecommand \bibfnamefont [1]{#1}%
\providecommand \citenamefont [1]{#1}%
\providecommand \href@noop [0]{\@secondoftwo}%
\providecommand \href [0]{\begingroup \@sanitize@url \@href}%
\providecommand \@href[1]{\@@startlink{#1}\@@href}%
\providecommand \@@href[1]{\endgroup#1\@@endlink}%
\providecommand \@sanitize@url [0]{\catcode `\\12\catcode `\$12\catcode `\&12\catcode `\#12\catcode `\^12\catcode `\_12\catcode `\%12\relax}%
\providecommand \@@startlink[1]{}%
\providecommand \@@endlink[0]{}%
\providecommand \url  [0]{\begingroup\@sanitize@url \@url }%
\providecommand \@url [1]{\endgroup\@href {#1}{\urlprefix }}%
\providecommand \urlprefix  [0]{URL }%
\providecommand \Eprint [0]{\href }%
\providecommand \doibase [0]{https://doi.org/}%
\providecommand \selectlanguage [0]{\@gobble}%
\providecommand \bibinfo  [0]{\@secondoftwo}%
\providecommand \bibfield  [0]{\@secondoftwo}%
\providecommand \translation [1]{[#1]}%
\providecommand \BibitemOpen [0]{}%
\providecommand \bibitemStop [0]{}%
\providecommand \bibitemNoStop [0]{.\EOS\space}%
\providecommand \EOS [0]{\spacefactor3000\relax}%
\providecommand \BibitemShut  [1]{\csname bibitem#1\endcsname}%
\let\auto@bib@innerbib\@empty
\bibitem [{\citenamefont {Becca}\ and\ \citenamefont {Sorella}(2017)}]{RN82}%
  \BibitemOpen
  \bibfield  {author} {\bibinfo {author} {\bibfnamefont {F.}~\bibnamefont {Becca}}\ and\ \bibinfo {author} {\bibfnamefont {S.}~\bibnamefont {Sorella}},\ }\href {https://doi.org/10.1017/9781316417041} {\bibinfo {title} {Quantum monte carlo approaches for correlated systems}} (\bibinfo {year} {2017})\BibitemShut {NoStop}%
\bibitem [{\citenamefont {Lloyd}(1996)}]{RN59}%
  \BibitemOpen
  \bibfield  {author} {\bibinfo {author} {\bibfnamefont {S.}~\bibnamefont {Lloyd}},\ }\bibfield  {title} {\bibinfo {title} {Universal quantum simulators},\ }\href {https://doi.org/10.1126/science.273.5278.1073} {\bibfield  {journal} {\bibinfo  {journal} {Science}\ }\textbf {\bibinfo {volume} {273}},\ \bibinfo {pages} {1073–1078} (\bibinfo {year} {1996})}\BibitemShut {NoStop}%
\bibitem [{\citenamefont {Georgescu}\ \emph {et~al.}(2014)\citenamefont {Georgescu}, \citenamefont {Ashhab},\ and\ \citenamefont {Nori}}]{RN60}%
  \BibitemOpen
  \bibfield  {author} {\bibinfo {author} {\bibfnamefont {I.~M.}\ \bibnamefont {Georgescu}}, \bibinfo {author} {\bibfnamefont {S.}~\bibnamefont {Ashhab}},\ and\ \bibinfo {author} {\bibfnamefont {F.}~\bibnamefont {Nori}},\ }\bibfield  {title} {\bibinfo {title} {Quantum simulation},\ }\href {https://doi.org/10.1103/RevModPhys.86.153} {\bibfield  {journal} {\bibinfo  {journal} {Rev Mod Phys}\ }\textbf {\bibinfo {volume} {86}},\ \bibinfo {pages} {153–185} (\bibinfo {year} {2014})}\BibitemShut {NoStop}%
\bibitem [{\citenamefont {Fauseweh}(2024)}]{RN72}%
  \BibitemOpen
  \bibfield  {author} {\bibinfo {author} {\bibfnamefont {B.}~\bibnamefont {Fauseweh}},\ }\bibfield  {title} {\bibinfo {title} {Quantum many-body simulations on digital quantum computers: State-of-the-art and future challenges},\ } \href {https://doi.org/10.1038/s41467-024-46402-9} {\bibfield  {journal} {\bibinfo  {journal} {Nat Comm}\ }\textbf {\bibinfo {volume} {15}},\  \bibinfo{pages} {2123}  (\bibinfo {year} {2024})}\BibitemShut {NoStop}%
\bibitem [{\citenamefont {Wilde}(2017)}]{RN65}%
  \BibitemOpen
  \bibfield  {author} {\bibinfo {author} {\bibfnamefont {M.~M.}\ \bibnamefont {Wilde}},\ }\href {https://doi.org/10.1017/9781316809976} {\bibinfo {title} {Quantum information theory}} (\bibinfo {year} {2017})\BibitemShut {NoStop}%
\bibitem [{\citenamefont {Holevo}(2012)}]{RN66}%
  \BibitemOpen
  \bibfield  {author} {\bibinfo {author} {\bibfnamefont {A.~S.}\ \bibnamefont {Holevo}},\ }\href {https://doi.org/10.1515/9783110273403} {\bibinfo {title} {Quantum systems, channels, information}} (\bibinfo {year} {2012})\BibitemShut {NoStop}%
\bibitem [{\citenamefont {Horodecki}\ \emph {et~al.}(2009)\citenamefont {Horodecki}, \citenamefont {Horodecki}, \citenamefont {Horodecki},\ and\ \citenamefont {Horodecki}}]{RN68}%
  \BibitemOpen
  \bibfield  {author} {\bibinfo {author} {\bibfnamefont {R.}~\bibnamefont {Horodecki}}, \bibinfo {author} {\bibfnamefont {P.}~\bibnamefont {Horodecki}}, \bibinfo {author} {\bibfnamefont {M.}~\bibnamefont {Horodecki}},\ and\ \bibinfo {author} {\bibfnamefont {K.}~\bibnamefont {Horodecki}},\ }\bibfield  {title} {\bibinfo {title} {Quantum entanglement},\ }\href {https://doi.org/10.1103/RevModPhys.81.865} {\bibfield  {journal} {\bibinfo  {journal} {Rev Mod Phys}\ }\textbf {\bibinfo {volume} {81}},\ \bibinfo {pages} {865–942} (\bibinfo {year} {2009})}\BibitemShut {NoStop}%
\bibitem [{\citenamefont {Amico}\ \emph {et~al.}(2008)\citenamefont {Amico}, \citenamefont {Fazio}, \citenamefont {Osterloh},\ and\ \citenamefont {Vedral}}]{RN73}%
  \BibitemOpen
  \bibfield  {author} {\bibinfo {author} {\bibfnamefont {L.}~\bibnamefont {Amico}}, \bibinfo {author} {\bibfnamefont {R.}~\bibnamefont {Fazio}}, \bibinfo {author} {\bibfnamefont {A.}~\bibnamefont {Osterloh}},\ and\ \bibinfo {author} {\bibfnamefont {V.}~\bibnamefont {Vedral}},\ }\bibfield  {title} {\bibinfo {title} {Entanglement in many-body systems},\ }\href {https://doi.org/10.1103/RevModPhys.80.517} {\bibfield  {journal} {\bibinfo  {journal} {Rev Mod Phys}\ }\textbf {\bibinfo {volume} {80}},\ \bibinfo {pages} {517–576} (\bibinfo {year} {2008})}\BibitemShut {NoStop}%
\bibitem [{\citenamefont {Reiher}\ \emph {et~al.}(2017)\citenamefont {Reiher}, \citenamefont {Wiebe}, \citenamefont {Svore}, \citenamefont {Wecker},\ and\ \citenamefont {Troyer}}]{RN69}%
  \BibitemOpen
  \bibfield  {author} {\bibinfo {author} {\bibfnamefont {M.}~\bibnamefont {Reiher}}, \bibinfo {author} {\bibfnamefont {N.}~\bibnamefont {Wiebe}}, \bibinfo {author} {\bibfnamefont {K.~M.}\ \bibnamefont {Svore}}, \bibinfo {author} {\bibfnamefont {D.}~\bibnamefont {Wecker}},\ and\ \bibinfo {author} {\bibfnamefont {M.}~\bibnamefont {Troyer}},\ }\bibfield  {title} {\bibinfo {title} {Elucidating reaction mechanisms on quantum computers},\ }\href {https://doi.org/10.1073/pnas.1619152114} {\bibfield  {journal} {\bibinfo  {journal} {PNAS}\ }\textbf {\bibinfo {volume} {114}},\ \bibinfo {pages} {7555–7560} (\bibinfo {year} {2017})}\BibitemShut {NoStop}%
\bibitem [{\citenamefont {Mcardle}\ \emph {et~al.}(2020)\citenamefont {Mcardle}, \citenamefont {Endo}, \citenamefont {Aspuru-Guzik}, \citenamefont {Benjamin},\ and\ \citenamefont {Yuan}}]{RN74}%
  \BibitemOpen
  \bibfield  {author} {\bibinfo {author} {\bibfnamefont {S.}~\bibnamefont {Mcardle}}, \bibinfo {author} {\bibfnamefont {S.}~\bibnamefont {Endo}}, \bibinfo {author} {\bibfnamefont {A.}~\bibnamefont {Aspuru-Guzik}}, \bibinfo {author} {\bibfnamefont {S.~C.}\ \bibnamefont {Benjamin}},\ and\ \bibinfo {author} {\bibfnamefont {X.}~\bibnamefont {Yuan}},\ }\bibfield  {title} {\bibinfo {title} {Quantum computational chemistry},\ }\href {https://doi.org/10.1103/RevModPhys.92.015003} {\bibfield  {journal} {\bibinfo  {journal} {Rev Mod Phys}\ }\textbf {\bibinfo {volume} {92}},\ \bibinfo {pages}  {015003} (\bibinfo {year} {2020})}\BibitemShut {NoStop}%
\bibitem [{\citenamefont {Jonathan}\ \emph {et~al.}(2017)\citenamefont {Jonathan}, \citenamefont {Yudong}, \citenamefont {Jonathan}, \citenamefont {Peter}, \citenamefont {Pierre-Luc}, \citenamefont {Nicolas}, \citenamefont {Prineha}, \citenamefont {Ian}, \citenamefont {Michael},\ and\ \citenamefont {Alán}}]{RN75}%
  \BibitemOpen
  \bibfield  {author} {\bibinfo {author} {\bibfnamefont {O.}~\bibnamefont {Jonathan}}, \bibinfo {author} {\bibfnamefont {C.}~\bibnamefont {Yudong}}, \bibinfo {author} {\bibfnamefont {R.}~\bibnamefont {Jonathan}}, \bibinfo {author} {\bibfnamefont {J.}~\bibnamefont {Peter}}, \bibinfo {author} {\bibfnamefont {D.-D.}\ \bibnamefont {Pierre-Luc}}, \bibinfo {author} {\bibfnamefont {S.}~\bibnamefont {Nicolas}}, \bibinfo {author} {\bibfnamefont {N.}~\bibnamefont {Prineha}}, \bibinfo {author} {\bibfnamefont {K.}~\bibnamefont {Ian}}, \bibinfo {author} {\bibfnamefont {W.}~\bibnamefont {Michael}},\ and\ \bibinfo {author} {\bibfnamefont {A.-G.}\ \bibnamefont {Alán}},\ } \bibfield {title} {\bibinfo {title} {Quantum information and computation for chemistry},\ }\href {https://doi.org/10.48550/arXiv.1706.05413} {\bibfield {journal} {\bibinfo {journal} {arXiv:1706.05413 [quant-ph]}} (\bibinfo {year} {2017})}\BibitemShut {NoStop}%
\bibitem [{\citenamefont {Sachdev}(2011)}]{RN76}%
  \BibitemOpen
  \bibfield  {author} {\bibinfo {author} {\bibfnamefont {S.}~\bibnamefont {Sachdev}},\ }\href {https://doi.org/10.1017/CBO9780511973765} {\bibinfo {title} {Quantum phase transitions}} (\bibinfo {year} {2011})\BibitemShut {NoStop}%
\bibitem [{\citenamefont {Anderson}(1972)}]{RN77}%
  \BibitemOpen
  \bibfield  {author} {\bibinfo {author} {\bibfnamefont {P.~W.}\ \bibnamefont {Anderson}},\ }\bibfield  {title} {\bibinfo {title} {More is different},\ }\href {https://doi.org/10.1126/science.177.4047.393} {\bibfield  {journal} {\bibinfo  {journal} {Science}\ }\textbf {\bibinfo {volume} {177}},\ \bibinfo {pages} {393–396} (\bibinfo {year} {1972})}\BibitemShut {NoStop}%
\bibitem [{\citenamefont {Schollwöck}(2011{\natexlab{a}})}]{RN79}%
  \BibitemOpen
  \bibfield  {author} {\bibinfo {author} {\bibfnamefont {U.}~\bibnamefont {Schollwöck}},\ }\bibfield  {title} {\bibinfo {title} {The density-matrix renormalization group in the age of matrix product states},\ }\href {https://doi.org/10.1016/j.aop.2010.09.012} {\bibfield  {journal} {\bibinfo  {journal} {Ann Phys}\ }\textbf {\bibinfo {volume} {326}},\ \bibinfo {pages} {96–192} (\bibinfo {year} {2011}{\natexlab{a}})}\BibitemShut {NoStop}%
\bibitem [{\citenamefont {Mcmillan}(1965)}]{RN83}%
  \BibitemOpen
  \bibfield  {author} {\bibinfo {author} {\bibfnamefont {W.~L.}\ \bibnamefont {McMillan}},\ }\bibfield  {title} {\bibinfo {title} {Ground state of liquid $\mathrm{He}^4$},\ }\href {https://doi.org/10.1103/PhysRev.138.A442} {\bibfield  {journal} {\bibinfo  {journal} {Phys Rev}\ }\textbf {\bibinfo {volume} {138}},\ \bibinfo {pages} {A442–A451} (\bibinfo {year} {1965})}\BibitemShut {NoStop}%
\bibitem [{\citenamefont {Ceperley}\ \emph {et~al.}(1977)\citenamefont {Ceperley}, \citenamefont {Chester},\ and\ \citenamefont {Kalos}}]{RN84}%
  \BibitemOpen
  \bibfield  {author} {\bibinfo {author} {\bibfnamefont {D.}~\bibnamefont {Ceperley}}, \bibinfo {author} {\bibfnamefont {G.~V.}\ \bibnamefont {Chester}},\ and\ \bibinfo {author} {\bibfnamefont {M.~H.}\ \bibnamefont {Kalos}},\ }\bibfield  {title} {\bibinfo {title} {Monte carlo simulation of a many-fermion study},\ }\href {https://doi.org/10.1103/PhysRevB.16.3081} {\bibfield  {journal} {\bibinfo  {journal} {Phys Rev B}\ }\textbf {\bibinfo {volume} {16}},\ \bibinfo {pages} {3081–3099} (\bibinfo {year} {1977})}\BibitemShut {NoStop}%
\bibitem [{\citenamefont {Foulkes}\ \emph {et~al.}(2001)\citenamefont {Foulkes}, \citenamefont {Mitas}, \citenamefont {Needs},\ and\ \citenamefont {Rajagopal}}]{RN85}%
  \BibitemOpen
  \bibfield  {author} {\bibinfo {author} {\bibfnamefont {W.~M.~C.}\ \bibnamefont {Foulkes}}, \bibinfo {author} {\bibfnamefont {L.}~\bibnamefont {Mitas}}, \bibinfo {author} {\bibfnamefont {R.~J.}\ \bibnamefont {Needs}},\ and\ \bibinfo {author} {\bibfnamefont {G.}~\bibnamefont {Rajagopal}},\ }\bibfield  {title} {\bibinfo {title} {Quantum monte carlo simulations of solids},\ }\href {https://doi.org/10.1103/RevModPhys.73.33} {\bibfield  {journal} {\bibinfo  {journal} {Rev Mod Phys}\ }\textbf {\bibinfo {volume} {73}},\ \bibinfo {pages} {33–83} (\bibinfo {year} {2001})}\BibitemShut {NoStop}%
\bibitem [{\citenamefont {Sorella}(2005)}]{RN86}%
  \BibitemOpen
  \bibfield  {author} {\bibinfo {author} {\bibfnamefont {S.}~\bibnamefont {Sorella}},\ }\bibfield  {title} {\bibinfo {title} {Wave function optimization in the variational monte carlo method},\ }\href {https://doi.org/10.1103/PhysRevB.71.241103} {\bibfield  {journal} {\bibinfo  {journal} {Phys Rev B}\ }\textbf {\bibinfo {volume} {71}},\  \bibinfo {pages} {241103} (\bibinfo {year} {2005})}\BibitemShut {NoStop}%
\bibitem [{\citenamefont {Sorella}(1998)}]{RN87}%
  \BibitemOpen
  \bibfield  {author} {\bibinfo {author} {\bibfnamefont {S.}~\bibnamefont {Sorella}},\ }\bibfield  {title} {\bibinfo {title} {Green function monte carlo with stochastic reconfiguration},\ }\href {https://doi.org/10.1103/PhysRevLett.80.4558} {\bibfield  {journal} {\bibinfo  {journal} {Phys Rev Lett}\ }\textbf {\bibinfo {volume} {80}},\ \bibinfo {pages} {4558–4561} (\bibinfo {year} {1998})}\BibitemShut {NoStop}%
\bibitem [{\citenamefont {Capriotti}\ \emph {et~al.}(2001)\citenamefont {Capriotti}, \citenamefont {Becca}, \citenamefont {Parola},\ and\ \citenamefont {Sorella}}]{RN88}%
  \BibitemOpen
  \bibfield  {author} {\bibinfo {author} {\bibfnamefont {L.}~\bibnamefont {Capriotti}}, \bibinfo {author} {\bibfnamefont {F.}~\bibnamefont {Becca}}, \bibinfo {author} {\bibfnamefont {A.}~\bibnamefont {Parola}},\ and\ \bibinfo {author} {\bibfnamefont {S.}~\bibnamefont {Sorella}},\ }\bibfield  {title} {\bibinfo {title} {Resonating valence bond wave functions for strongly frustrated spin systems},\ }\href {https://doi.org/10.1103/PhysRevLett.87.097201} {\bibfield  {journal} {\bibinfo  {journal} {Phys Rev Lett}\ }\textbf {\bibinfo {volume} {87}},\ \bibinfo {pages} {097201} (\bibinfo {year} {2001})}\BibitemShut {NoStop}%
\bibitem [{\citenamefont {Jastrow}(1955)}]{RN39}%
  \BibitemOpen
  \bibfield  {author} {\bibinfo {author} {\bibfnamefont {R.}~\bibnamefont {Jastrow}},\ }\bibfield  {title} {\bibinfo {title} {Many-body problem with strong forces},\ }\href {https://doi.org/10.1103/PhysRev.98.1479} {\bibfield  {journal} {\bibinfo  {journal} {Phys Rev}\ }\textbf {\bibinfo {volume} {98}},\ \bibinfo {pages} {1479–1484} (\bibinfo {year} {1955})}\BibitemShut {NoStop}%
\bibitem [{\citenamefont {Gutzwiller}(1963)}]{RN40}%
  \BibitemOpen
  \bibfield  {author} {\bibinfo {author} {\bibfnamefont {M.~C.}\ \bibnamefont {Gutzwiller}},\ }\bibfield  {title} {\bibinfo {title} {Effect of correlation on the ferromagnetism of transition metals},\ }\href {https://doi.org/10.1103/PhysRevLett.10.159} {\bibfield  {journal} {\bibinfo  {journal} {Phys Rev Lett}\ }\textbf {\bibinfo {volume} {10}},\ \bibinfo {pages} {159–162} (\bibinfo {year} {1963})}\BibitemShut {NoStop}%
\bibitem [{\citenamefont {Schollwöck}(2011{\natexlab{b}})}]{DMRG}%
  \BibitemOpen
  \bibfield  {author} {\bibinfo {author} {\bibfnamefont {U.}~\bibnamefont {Schollwöck}},\ }\bibfield  {title} {\bibinfo {title} {The density-matrix renormalization group in the age of matrix product states},\ }\href {https://doi.org/https://doi.org/10.1016/j.aop.2010.09.012} {\bibfield  {journal} {\bibinfo  {journal} {Ann Phys}\ }\textbf {\bibinfo {volume} {326}},\ \bibinfo {pages} {96} (\bibinfo {year} {2011}{\natexlab{b}})}\BibitemShut {NoStop}%
\bibitem [{\citenamefont {Carleo}\ and\ \citenamefont {Troyer}(2017{\natexlab{a}})}]{RN80}%
  \BibitemOpen
  \bibfield  {author} {\bibinfo {author} {\bibfnamefont {G.}~\bibnamefont {Carleo}}\ and\ \bibinfo {author} {\bibfnamefont {M.}~\bibnamefont {Troyer}},\ }\bibfield  {title} {\bibinfo {title} {Solving the quantum many-body problem with artificial neural networks},\ }\href {https://doi.org/10.1126/science.aag2302} {\bibfield  {journal} {\bibinfo  {journal} {Science}\ }\textbf {\bibinfo {volume} {355}},\ \bibinfo {pages} {602–606} (\bibinfo {year} {2017}{\natexlab{a}})}\BibitemShut {NoStop}%
\bibitem [{\citenamefont {Sharir}\ \emph {et~al.}(2020)\citenamefont {Sharir}, \citenamefont {Levine}, \citenamefont {Wies}, \citenamefont {Carleo},\ and\ \citenamefont {Shashua}}]{SharirEtAl}%
  \BibitemOpen
  \bibfield  {author} {\bibinfo {author} {\bibfnamefont {O.}~\bibnamefont {Sharir}}, \bibinfo {author} {\bibfnamefont {Y.}~\bibnamefont {Levine}}, \bibinfo {author} {\bibfnamefont {N.}~\bibnamefont {Wies}}, \bibinfo {author} {\bibfnamefont {G.}~\bibnamefont {Carleo}},\ and\ \bibinfo {author} {\bibfnamefont {A.}~\bibnamefont {Shashua}},\ }\bibfield  {title} {\bibinfo {title} {Deep autoregressive models for the efficient variational simulation of many-body quantum systems},\ }\href {https://doi.org/10.1103/PhysRevLett.124.020503} {\bibfield  {journal} {\bibinfo  {journal} {Phys Rev Lett}\ }\textbf {\bibinfo {volume} {124}},\ \bibinfo {pages} {020503} (\bibinfo {year} {2020})}\BibitemShut {NoStop}%
\bibitem [{\citenamefont {Lange}\ \emph {et~al.}(2024)\citenamefont {Lange}, \citenamefont {Van~de Walle}, \citenamefont {Abedinnia},\ and\ \citenamefont {Bohrdt}}]{Lange_2024}%
  \BibitemOpen
  \bibfield  {author} {\bibinfo {author} {\bibfnamefont {H.}~\bibnamefont {Lange}}, \bibinfo {author} {\bibfnamefont {A.}~\bibnamefont {Van~de Walle}}, \bibinfo {author} {\bibfnamefont {A.}~\bibnamefont {Abedinnia}},\ and\ \bibinfo {author} {\bibfnamefont {A.}~\bibnamefont {Bohrdt}},\ }\bibfield  {title} {\bibinfo {title} {From architectures to applications: a review of neural quantum states},\ }\href {https://doi.org/10.1088/2058-9565/ad7168} {\bibfield  {journal} {\bibinfo  {journal} {Quantum Sci Technol}\ }\textbf {\bibinfo {volume} {9}},\ \bibinfo {pages} {040501} (\bibinfo {year} {2024})}\BibitemShut {NoStop}%
\bibitem [{\citenamefont {Jia}\ \emph {et~al.}(2019)\citenamefont {Jia}, \citenamefont {Yi}, \citenamefont {Zhai}, \citenamefont {Wu}, \citenamefont {Guo},\ and\ \citenamefont {Guo}}]{Jia_2019}%
  \BibitemOpen
  \bibfield  {author} {\bibinfo {author} {\bibfnamefont {Z.}~\bibnamefont {Jia}}, \bibinfo {author} {\bibfnamefont {B.}~\bibnamefont {Yi}}, \bibinfo {author} {\bibfnamefont {R.}~\bibnamefont {Zhai}}, \bibinfo {author} {\bibfnamefont {Y.}~\bibnamefont {Wu}}, \bibinfo {author} {\bibfnamefont {G.}~\bibnamefont {Guo}},\ and\ \bibinfo {author} {\bibfnamefont {G.}~\bibnamefont {Guo}},\ }\bibfield  {title} {\bibinfo {title} {Quantum neural network states: A brief review of methods and applications},\ }\href {https://doi.org/10.1002/qute.201800077} {\bibfield  {journal} {\bibinfo  {journal} {Adv Quantum Technol}\ }\textbf {\bibinfo {volume} {2}},\ \bibfield {pages} {1800077} (\bibinfo {year} {2019})}\BibitemShut {NoStop}%
\bibitem [{\citenamefont {Hibat-Allah}\ \emph {et~al.}(2020)\citenamefont {Hibat-Allah}, \citenamefont {Ganahl}, \citenamefont {Hayward}, \citenamefont {Melko},\ and\ \citenamefont {Carrasquilla}}]{Hibat-a}%
  \BibitemOpen
  \bibfield  {author} {\bibinfo {author} {\bibfnamefont {M.}~\bibnamefont {Hibat-Allah}}, \bibinfo {author} {\bibfnamefont {M.}~\bibnamefont {Ganahl}}, \bibinfo {author} {\bibfnamefont {L.~E.}\ \bibnamefont {Hayward}}, \bibinfo {author} {\bibfnamefont {R.~G.}\ \bibnamefont {Melko}},\ and\ \bibinfo {author} {\bibfnamefont {J.}~\bibnamefont {Carrasquilla}},\ }\bibfield  {title} {\bibinfo {title} {Recurrent neural network wave functions},\ }\href {https://doi.org/10.1103/PhysRevResearch.2.023358} {\bibfield  {journal} {\bibinfo  {journal} {Phys Rev Res}\ }\textbf {\bibinfo {volume} {2}},\ \bibinfo {pages} {023358} (\bibinfo {year} {2020})}\BibitemShut {NoStop}%
\bibitem [{\citenamefont {Hibat-Allah}\ \emph {et~al.}(2025)\citenamefont {Hibat-Allah}, \citenamefont {Merali}, \citenamefont {Torlai}, \citenamefont {Melko},\ and\ \citenamefont {Carrasquilla}}]{Hibat-b}%
  \BibitemOpen
  \bibfield  {author} {\bibinfo {author} {\bibfnamefont {M.}~\bibnamefont {Hibat-Allah}}, \bibinfo {author} {\bibfnamefont {E.}~\bibnamefont {Merali}}, \bibinfo {author} {\bibfnamefont {G.}~\bibnamefont {Torlai}}, \bibinfo {author} {\bibfnamefont {R.~G.}\ \bibnamefont {Melko}},\ and\ \bibinfo {author} {\bibfnamefont {J.}~\bibnamefont {Carrasquilla}},\ }\bibfield  {title} {\bibinfo {title} {Recurrent neural network wave functions for rydberg atom arrays on kagome lattice},\ }\href {https://doi.org/10.1038/s42005-025-02226-7} {\bibfield  {journal} {\bibinfo  {journal} {Commun Phys}\ }\textbf {\bibinfo {volume} {3}} (\bibinfo {year} {2025})}\BibitemShut {NoStop}%
\bibitem [{\citenamefont {Vaswani}\ \emph {et~al.}(2017)\citenamefont {Vaswani}, \citenamefont {Shazeer}, \citenamefont {Parmar}, \citenamefont {Uszkoreit}, \citenamefont {Jones}, \citenamefont {Gomez}, \citenamefont {Kaiser},\ and\ \citenamefont {Polosukhin}}]{vaswani2017attention}%
  \BibitemOpen
  \bibfield  {author} {\bibinfo {author} {\bibfnamefont {A.}~\bibnamefont {Vaswani}}, \bibinfo {author} {\bibfnamefont {N.}~\bibnamefont {Shazeer}}, \bibinfo {author} {\bibfnamefont {N.}~\bibnamefont {Parmar}}, \bibinfo {author} {\bibfnamefont {J.}~\bibnamefont {Uszkoreit}}, \bibinfo {author} {\bibfnamefont {L.}~\bibnamefont {Jones}}, \bibinfo {author} {\bibfnamefont {A.~N.}\ \bibnamefont {Gomez}}, \bibinfo {author} {\bibfnamefont {{\L}.}~\bibnamefont {Kaiser}},\ and\ \bibinfo {author} {\bibfnamefont {I.}~\bibnamefont {Polosukhin}},\ }\bibfield  {title} {\bibinfo {title} {Attention is all you need},\ }in\ \href {https://doi.org/10.48550/arXiv.1706.03762} {\bibfield {journal} {\bibinfo {journal} {arXiv:1706.03762 [cs.CL]}\ } (\bibinfo {year} {2017})}\BibitemShut {NoStop}%
\bibitem [{\citenamefont {Sprague}\ and\ \citenamefont {Czischek}(2024)}]{Sprague&Czischek}%
  \BibitemOpen
  \bibfield  {author} {\bibinfo {author} {\bibfnamefont {K.}~\bibnamefont {Sprague}}\ and\ \bibinfo {author} {\bibfnamefont {S.}~\bibnamefont {Czischek}},\ }\bibfield  {title} {\bibinfo {title} {Variational monte carlo with large patched transformers},\ } \href {https://doi.org/10.1038/s42005-024-01584-y} {\bibfield  {journal} {\bibinfo  {journal} {Communications Physics}\ }\textbf {\bibinfo {volume} {7}},\ \bibfield {pages} {90} (\bibinfo {year} {2024})}\BibitemShut {NoStop}%
\bibitem [{\citenamefont {Wu}\ \emph {et~al.}(2023)\citenamefont {Wu}, \citenamefont {Guo}, \citenamefont {Fan}, \citenamefont {Zhou},\ and\ \citenamefont {Shang}}]{RN142}%
  \BibitemOpen
  \bibfield  {author} {\bibinfo {author} {\bibfnamefont {Y.}~\bibnamefont {Wu}}, \bibinfo {author} {\bibfnamefont {C.}~\bibnamefont {Guo}}, \bibinfo {author} {\bibfnamefont {Y.}~\bibnamefont {Fan}}, \bibinfo {author} {\bibfnamefont {P.}~\bibnamefont {Zhou}},\ and\ \bibinfo {author} {\bibfnamefont {H.}~\bibnamefont {Shang}},\ }\href {https://doi.org/10.1145/3581784.3607061} {\bibinfo {title} {NNQS-transformer: an efficient and scalable neural network quantum states approach for ab initio quantum chemistry}} (\bibinfo {year} {2023})\BibitemShut {NoStop}%
\bibitem [{\citenamefont {Trockman}\ and\ \citenamefont {Kolter}(2022)}]{trockman2022patches}%
  \BibitemOpen
  \bibfield  {author} {\bibinfo {author} {\bibfnamefont {A.}~\bibnamefont {Trockman}}\ and\ \bibinfo {author} {\bibfnamefont {J.~Z.}\ \bibnamefont {Kolter}},\ } \bibfield {title} {\bibinfo {title} {Patches are all you need?,\ }}\href {https://doi.org/10.48550/arXiv.2201.09792} {\bibfield {journal} {\bibinfo {journal} {arXiv:2201.09792[cs.CV]}} (\bibinfo {year} {2022})}\BibitemShut {NoStop}%
\bibitem [{\citenamefont {Viteritti}\ \emph {et~al.}(2025)\citenamefont {Viteritti}, \citenamefont {Rende}, \citenamefont {Parola}, \citenamefont {Goldt},\ and\ \citenamefont {Becca}}]{Viteritti}%
  \BibitemOpen
  \bibfield  {author} {\bibinfo {author} {\bibfnamefont {L.~L.}\ \bibnamefont {Viteritti}}, \bibinfo {author} {\bibfnamefont {R.}~\bibnamefont {Rende}}, \bibinfo {author} {\bibfnamefont {A.}~\bibnamefont {Parola}}, \bibinfo {author} {\bibfnamefont {S.}~\bibnamefont {Goldt}},\ and\ \bibinfo {author} {\bibfnamefont {F.}~\bibnamefont {Becca}},\ }\bibfield  {title} {\bibinfo {title} {Transformer wave function for two dimensional frustrated magnets: Emergence of a spin-liquid phase in the shastry-sutherland model},\ }\href {https://doi.org/10.1103/PhysRevB.111.134411} {\bibfield  {journal} {\bibinfo  {journal} {Phys Rev B}\ }\textbf {\bibinfo {volume} {111}},\ \bibinfo {pages} {134411} (\bibinfo {year} {2025})}\BibitemShut {NoStop}%
\bibitem [{\citenamefont {Rende}\ and\ \citenamefont {Viteritti}(2025)}]{RN143}%
  \BibitemOpen
  \bibfield  {author} {\bibinfo {author} {\bibfnamefont {R.}~\bibnamefont {Rende}}\ and\ \bibinfo {author} {\bibfnamefont {L.~L.}\ \bibnamefont {Viteritti}},\ }\bibfield  {title} {\bibinfo {title} {Are queries and keys always relevant? a case study on transformer wave functions},\ }\href {https://doi.org/10.1088/2632-2153/ada1a0} {\bibfield  {journal} {\bibinfo  {journal} {Mach Learn: Scie Technol}\ }\textbf {\bibinfo {volume} {6}},\ \bibinfo {pages} {010501} (\bibinfo {year} {2025})}\BibitemShut {NoStop}%
\bibitem [{\citenamefont {Tolstikhin}\ \emph {et~al.}(2021)\citenamefont {Tolstikhin}, \citenamefont {Houlsby}, \citenamefont {Kolesnikov}, \citenamefont {Beyer}, \citenamefont {Zhai}, \citenamefont {Unterthiner}, \citenamefont {Yung}, \citenamefont {Steiner}, \citenamefont {Keysers}, \citenamefont {Uszkoreit}, \citenamefont {Lucic},\ and\ \citenamefont {Dosovitskiy}}]{tolstikhin2021mlpmixer}%
  \BibitemOpen
  \bibfield  {author} {\bibinfo {author} {\bibfnamefont {I.}~\bibnamefont {Tolstikhin}}, \bibinfo {author} {\bibfnamefont {N.}~\bibnamefont {Houlsby}}, \bibinfo {author} {\bibfnamefont {A.}~\bibnamefont {Kolesnikov}}, \bibinfo {author} {\bibfnamefont {L.}~\bibnamefont {Beyer}}, \bibinfo {author} {\bibfnamefont {X.}~\bibnamefont {Zhai}}, \bibinfo {author} {\bibfnamefont {T.}~\bibnamefont {Unterthiner}}, \bibinfo {author} {\bibfnamefont {J.}~\bibnamefont {Yung}}, \bibinfo {author} {\bibfnamefont {A.~P.}\ \bibnamefont {Steiner}}, \bibinfo {author} {\bibfnamefont {D.}~\bibnamefont {Keysers}}, \bibinfo {author} {\bibfnamefont {J.}~\bibnamefont {Uszkoreit}}, \bibinfo {author} {\bibfnamefont {M.}~\bibnamefont {Lucic}},\ and\ \bibinfo {author} {\bibfnamefont {A.}~\bibnamefont {Dosovitskiy}},\ }\bibfield  {title} {\bibinfo {title} {{MLP}-mixer: An all-{MLP} architecture for vision},\ }in\ \href {https://openreview.net/forum?id=EI2KOXKdnP} {\emph {\bibinfo {booktitle} {Adv Neural Inf Process
  Systems}}},\ \bibinfo {editor} {edited by\ \bibinfo {editor} {\bibfnamefont {A.}~\bibnamefont {Beygelzimer}}, \bibinfo {editor} {\bibfnamefont {Y.}~\bibnamefont {Dauphin}}, \bibinfo {editor} {\bibfnamefont {P.}~\bibnamefont {Liang}},\ and\ \bibinfo {editor} {\bibfnamefont {J.~W.}\ \bibnamefont {Vaughan}}}\ (\bibinfo {year} {2021})\BibitemShut {NoStop}%
\bibitem [{\citenamefont {Cho}\ \emph {et~al.}(2014{\natexlab{a}})\citenamefont {Cho}, \citenamefont {van Merri{\"e}nboer}, \citenamefont {Gulcehre}, \citenamefont {Bahdanau}, \citenamefont {Bougares}, \citenamefont {Schwenk},\ and\ \citenamefont {Bengio}}]{cho-etal-2014-learning}%
  \BibitemOpen
  \bibfield  {author} {\bibinfo {author} {\bibfnamefont {K.}~\bibnamefont {Cho}}, \bibinfo {author} {\bibfnamefont {B.}~\bibnamefont {van Merri{\"e}nboer}}, \bibinfo {author} {\bibfnamefont {C.}~\bibnamefont {Gulcehre}}, \bibinfo {author} {\bibfnamefont {D.}~\bibnamefont {Bahdanau}}, \bibinfo {author} {\bibfnamefont {F.}~\bibnamefont {Bougares}}, \bibinfo {author} {\bibfnamefont {H.}~\bibnamefont {Schwenk}},\ and\ \bibinfo {author} {\bibfnamefont {Y.}~\bibnamefont {Bengio}},\ }\bibfield  {title} {\bibinfo {title} {Learning phrase representations using {RNN} encoder{--}decoder for statistical machine translation},\ }in\ \href {https://doi.org/10.3115/v1/D14-1179} {\emph {\bibinfo {booktitle} {Proceedings of EMNLP 2014}}},\ \bibinfo {editor} {edited by\ \bibinfo {editor} {\bibfnamefont {A.}~\bibnamefont {Moschitti}}, \bibinfo {editor} {\bibfnamefont {B.}~\bibnamefont {Pang}},\ and\ \bibinfo {editor} {\bibfnamefont {W.}~\bibnamefont
  {Daelemans}}}\ (\bibinfo  {publisher} {Association for Computational Linguistics},\ \bibinfo {address} {Doha, Qatar},\ \bibinfo {year} {2014})\ pp.\ \bibinfo {pages} {1724--1734} (\bibinfo {year} {2014})\BibitemShut {NoStop}%
\bibitem [{\citenamefont {van~den Oord}\ \emph {et~al.}(2016)\citenamefont {van~den Oord}, \citenamefont {Kalchbrenner},\ and\ \citenamefont {Kavukcuoglu}}]{PixelRNN}%
  \BibitemOpen
  \bibfield  {author} {\bibinfo {author} {\bibfnamefont {A.}~\bibnamefont {van~den Oord}}, \bibinfo {author} {\bibfnamefont {N.}~\bibnamefont {Kalchbrenner}},\ and\ \bibinfo {author} {\bibfnamefont {K.}~\bibnamefont {Kavukcuoglu}},\ }\bibfield  {title} {\bibinfo {title} {Pixel recurrent neural networks},\ }\href {https://doi.org/10.48550/arXiv.1601.06759} {\bibfield  {journal} {\bibinfo  {journal} {arXiv:1601.06759[cs.CV]}\ } (\bibinfo {year} {2016})}\BibitemShut {NoStop}%
\bibitem [{\citenamefont {Teoh}\ and\ \citenamefont {Melko}(2024)}]{teoh2024autoregressivemodelpathdependence}%
  \BibitemOpen
  \bibfield  {author} {\bibinfo {author} {\bibfnamefont {Y.~H.}\ \bibnamefont {Teoh}}\ and\ \bibinfo {author} {\bibfnamefont {R.~G.}\ \bibnamefont {Melko}},\ } \bibfield{title}{\bibinfo {title} {Autoregressive model path dependence near ising criticality}} \href {https://doi.org/10.48550/arXiv.2408.15715} {\bibfield {journal} {\bibinfo {journal} {arXiv:2408.15715[cs.LG]}} (\bibinfo {year} {2024})} \BibitemShut {NoStop}%
\bibitem [{\citenamefont {Moss}\ \emph {et~al.}(2025)\citenamefont {Moss}, \citenamefont {Wiersema}, \citenamefont {Hibat-Allah}, \citenamefont {Carrasquilla},\ and\ \citenamefont {Melko}}]{RN140}%
  \BibitemOpen
  \bibfield  {author} {\bibinfo {author} {\bibfnamefont {M.~S.}\ \bibnamefont {Moss}}, \bibinfo {author} {\bibfnamefont {R.}~\bibnamefont {Wiersema}}, \bibinfo {author} {\bibfnamefont {M.}~\bibnamefont {Hibat-Allah}}, \bibinfo {author} {\bibfnamefont {J.}~\bibnamefont {Carrasquilla}},\ and\ \bibinfo {author} {\bibfnamefont {R.~G.}\ \bibnamefont {Melko}},\ }\bibfield  {title} {\bibinfo {title} {Leveraging recurrence in neural network wavefunctions for large-scale simulations of heisenberg antiferromagnets on the square lattice},\ } \href {https://doi.org/10.1103/6ccd-wzhz}{ \bibfield  {journal} {\bibinfo  {journal} {Phys Rev B}\ }\textbf {\bibinfo {volume} {112}},\ \bibinfo{pages} {134450} (\bibinfo {year} {2025})}\BibitemShut {NoStop}%
\bibitem [{\citenamefont {Bengio}\ and\ \citenamefont {Bengio}(1999)}]{NIPS1999_e6384711}%
  \BibitemOpen
  \bibfield  {author} {\bibinfo {author} {\bibfnamefont {Y.}~\bibnamefont {Bengio}}\ and\ \bibinfo {author} {\bibfnamefont {S.}~\bibnamefont {Bengio}},\ }\bibfield  {title} {\bibinfo {title} {Modeling high-dimensional discrete data with multi-layer neural networks},\ }in\ \href {https://proceedings.neurips.cc/paper_files/paper/1999/file/e6384711491713d29bc63fc5eeb5ba4f-Paper.pdf} {\emph {\bibinfo {booktitle} {Adv Neural Inf Process Syst}}},\ Vol.~\bibinfo {volume} {12},\ \bibinfo {editor} {edited by\ \bibinfo {editor} {\bibfnamefont {S.}~\bibnamefont {Solla}}, \bibinfo {editor} {\bibfnamefont {T.}~\bibnamefont {Leen}},\ and\ \bibinfo {editor} {\bibfnamefont {K.}~\bibnamefont {M\"{u}ller}}}\ (\bibinfo  {publisher} {MIT Press},\ \bibinfo {year} {1999})\BibitemShut {NoStop}%
\bibitem [{\citenamefont {Germain}\ \emph {et~al.}(2015{\natexlab{a}})\citenamefont {Germain}, \citenamefont {Gregor}, \citenamefont {Murray},\ and\ \citenamefont {Larochelle}}]{germain2015made}%
  \BibitemOpen
  \bibfield  {author} {\bibinfo {author} {\bibfnamefont {M.}~\bibnamefont {Germain}}, \bibinfo {author} {\bibfnamefont {K.}~\bibnamefont {Gregor}}, \bibinfo {author} {\bibfnamefont {I.}~\bibnamefont {Murray}},\ and\ \bibinfo {author} {\bibfnamefont {H.}~\bibnamefont {Larochelle}},\ }\bibfield  {title} {\bibinfo {title} {Made: Masked autoencoder for distribution estimation},\ }in\ \href {https://doi.org/10.48550/arXiv.1502.03509} {\bibfield {journal} {\bibinfo {journal} {arXiv:1502.03509[cs.LG]}}\ (\bibinfo {year} {2015})}\ \BibitemShut {NoStop}%
\bibitem [{\citenamefont {Hibat-Allah}\ \emph {et~al.}(2022)\citenamefont {Hibat-Allah}, \citenamefont {Melko},\ and\ \citenamefont {Carrasquilla}}]{HibatAllah2022SupplementingRN}%
  \BibitemOpen
  \bibfield  {author} {\bibinfo {author} {\bibfnamefont {M.}~\bibnamefont {Hibat-Allah}}, \bibinfo {author} {\bibfnamefont {R.~G.}\ \bibnamefont {Melko}},\ and\ \bibinfo {author} {\bibfnamefont {J.}~\bibnamefont {Carrasquilla}},\ }\bibfield  {title} {\bibinfo {title} {{Supplementing Recurrent Neural Network Wave Functions with Symmetry and Annealing to Improve Accuracy}},\ }in\ \href {https://doi.org/10.48550/arXiv.2207.14314} {\bibfield {journal} {\bibinfo {journal} {arXiv:2207.14314[cond-mat.dis-nn]}}\ (\bibinfo {year} {2022})} \BibitemShut {NoStop}%
\bibitem [{\citenamefont {Carleo}\ and\ \citenamefont {Troyer}(2017{\natexlab{b}})}]{CarleoandTroyer}%
  \BibitemOpen
  \bibfield  {author} {\bibinfo {author} {\bibfnamefont {G.}~\bibnamefont {Carleo}}\ and\ \bibinfo {author} {\bibfnamefont {M.}~\bibnamefont {Troyer}},\ }\bibfield  {title} {\bibinfo {title} {Solving the quantum many-body problem with artificial neural networks},\ }\href {https://doi.org/10.1126/science.aag2302} {\bibfield  {journal} {\bibinfo  {journal} {Science}\ }\textbf {\bibinfo {volume} {355}},\ \bibinfo {pages} {602} (\bibinfo {year} {2017})} \BibitemShut {NoStop}%
\bibitem [{\citenamefont {Elman}(1990)}]{RN145}%
  \BibitemOpen
  \bibfield  {author} {\bibinfo {author} {\bibfnamefont {J.~L.}\ \bibnamefont {Elman}},\ }\bibfield  {title} {\bibinfo {title} {Finding structure in time},\ }\href {https://doi.org/10.1207/s15516709cog1402_1} {\bibfield  {journal} {\bibinfo  {journal} {Cog Sci}\ }\textbf {\bibinfo {volume} {14}},\ \bibinfo {pages} {179–211} (\bibinfo {year} {1990})}\BibitemShut {NoStop}%
\bibitem [{\citenamefont {Cho}\ \emph {et~al.}(2014{\natexlab{b}})\citenamefont {Cho}, \citenamefont {van Merri{\"e}nboer}, \citenamefont {Bahdanau},\ and\ \citenamefont {Bengio}}]{cho-etal-2014-properties}%
  \BibitemOpen
  \bibfield  {author} {\bibinfo {author} {\bibfnamefont {K.}~\bibnamefont {Cho}}, \bibinfo {author} {\bibfnamefont {B.}~\bibnamefont {van Merri{\"e}nboer}}, \bibinfo {author} {\bibfnamefont {D.}~\bibnamefont {Bahdanau}},\ and\ \bibinfo {author} {\bibfnamefont {Y.}~\bibnamefont {Bengio}},\ }\bibfield  {title} {\bibinfo {title} {On the properties of neural machine translation: Encoder{--}decoder approaches},\ }in\ \href {https://doi.org/10.3115/v1/W14-4012} {\emph {\bibinfo {booktitle} {Proceedings of {SSST}-8}}},\ \bibinfo {editor} {edited by\ \bibinfo {editor} {\bibfnamefont {D.}~\bibnamefont {Wu}}, \bibinfo {editor} {\bibfnamefont {M.}~\bibnamefont {Carpuat}}, \bibinfo {editor} {\bibfnamefont {X.}~\bibnamefont {Carreras}},\ and\ \bibinfo {editor} {\bibfnamefont {E.~M.}\ \bibnamefont {Vecchi}}}\ (\bibinfo  {publisher} {Association for Computational Linguistics},\ \bibinfo {address} {Doha, Qatar},\ \bibinfo {year} {2014})\ pp.\
  \bibinfo {pages} {103--111}\BibitemShut {NoStop}%
\bibitem [{\citenamefont {Cha}\ \emph {et~al.}(2021)\citenamefont {Cha}, \citenamefont {Ginsparg}, \citenamefont {Wu}, \citenamefont {Carrasquilla}, \citenamefont {McMahon},\ and\ \citenamefont {Kim}}]{Cha_2022}%
  \BibitemOpen
  \bibfield  {author} {\bibinfo {author} {\bibfnamefont {P.}~\bibnamefont {Cha}}, \bibinfo {author} {\bibfnamefont {P.}~\bibnamefont {Ginsparg}}, \bibinfo {author} {\bibfnamefont {F.}~\bibnamefont {Wu}}, \bibinfo {author} {\bibfnamefont {J.}~\bibnamefont {Carrasquilla}}, \bibinfo {author} {\bibfnamefont {P.~L.}\ \bibnamefont {McMahon}},\ and\ \bibinfo {author} {\bibfnamefont {E.-A.}\ \bibnamefont {Kim}},\ }\bibfield  {title} {\bibinfo {title} {Attention-based quantum tomography},\ }\href {https://doi.org/10.1088/2632-2153/ac362b} {\bibfield  {journal} {\bibinfo  {journal} {Mach Learn: Sci Technol}\ }\textbf {\bibinfo {volume} {3}},\ \bibinfo {pages} {01LT01} (\bibinfo {year} {2021})}\BibitemShut {NoStop}%
\bibitem [{\citenamefont {Bin}\ \emph {et~al.}()\citenamefont {Bin}, \citenamefont {Mohamed},\ and\ \citenamefont {Mohamed}}]{RN134}%
  \BibitemOpen
  \bibfield  {author} {\bibinfo {author} {\bibfnamefont {A.~A.}\ \bibnamefont {Bin}}, \bibinfo {author} {\bibfnamefont {A.~A.}\ \bibnamefont {Mohamed}},\ and\ \bibinfo {author} {\bibfnamefont {H.-A.}\ \bibnamefont {Mohamed}},\ } \bibinfo {title} {Geometry-induced long-range correlations in recurrent neural network quantum states,\ }\href {https://doi.org/10.48550/arXiv.2604.08661} {\bibfield {journal} {\bibinfo {journal} {arXiv:2604.08661[quant-ph]}} (\bibinfo {year} {2026})}\BibitemShut {NoStop}%
\bibitem [{\citenamefont {Ibarra-Garc\'{\i}a-Padilla}\ \emph {et~al.}(2025)\citenamefont {Ibarra-Garc\'{\i}a-Padilla}, \citenamefont {Lange}, \citenamefont {Melko}, \citenamefont {Scalettar}, \citenamefont {Carrasquilla}, \citenamefont {Bohrdt},\ and\ \citenamefont {Khatami}}]{PhysRevResearch.7.013122}%
  \BibitemOpen
  \bibfield  {author} {\bibinfo {author} {\bibfnamefont {E.}~\bibnamefont {Ibarra-Garc\'{\i}a-Padilla}}, \bibinfo {author} {\bibfnamefont {H.}~\bibnamefont {Lange}}, \bibinfo {author} {\bibfnamefont {R.~G.}\ \bibnamefont {Melko}}, \bibinfo {author} {\bibfnamefont {R.~T.}\ \bibnamefont {Scalettar}}, \bibinfo {author} {\bibfnamefont {J.}~\bibnamefont {Carrasquilla}}, \bibinfo {author} {\bibfnamefont {A.}~\bibnamefont {Bohrdt}},\ and\ \bibinfo {author} {\bibfnamefont {E.}~\bibnamefont {Khatami}},\ }\bibfield  {title} {\bibinfo {title} {Autoregressive neural quantum states of fermi hubbard models},\ }\href {https://doi.org/10.1103/PhysRevResearch.7.013122} {\bibfield  {journal} {\bibinfo  {journal} {Phys Rev Res}\ }\textbf {\bibinfo {volume} {7}},\ \bibinfo {pages} {013122} (\bibinfo {year} {2025})}\BibitemShut {NoStop}%
\bibitem [{\citenamefont {Ayub}\ \emph {et~al.}(2026)\citenamefont {Ayub}, \citenamefont {Aboussalah},\ and\ \citenamefont {Hibat-Allah}}]{ayub2026geometryinducedlongrangecorrelationsrecurrent}%
  \BibitemOpen
  \bibfield  {author} {\bibinfo {author} {\bibfnamefont {A.~B.}\ \bibnamefont {Ayub}}, \bibinfo {author} {\bibfnamefont {A.~M.}\ \bibnamefont {Aboussalah}},\ and\ \bibinfo {author} {\bibfnamefont {M.}~\bibnamefont {Hibat-Allah}},\ } \bibinfo {title} {Geometry-induced long-range correlations in recurrent neural network quantum states,\ }\href {https://doi.org/10.48550/arXiv.2604.08661} {\bibfield{journal} {\bibinfo{journal} {arXiv:2604.08661[quant-ph]}} (\bibinfo {year} {2026})}\BibitemShut {NoStop}%
\bibitem [{\citenamefont {Viteritti}\ \emph {et~al.}(2023)\citenamefont {Viteritti}, \citenamefont {Rende},\ and\ \citenamefont {Becca}}]{Luciano}%
  \BibitemOpen
  \bibfield  {author} {\bibinfo {author} {\bibfnamefont {L.~L.}\ \bibnamefont {Viteritti}}, \bibinfo {author} {\bibfnamefont {R.}~\bibnamefont {Rende}},\ and\ \bibinfo {author} {\bibfnamefont {F.}~\bibnamefont {Becca}},\ }\bibfield  {title} {\bibinfo {title} {Transformer variational wave functions for frustrated quantum spin systems},\ }\href {https://doi.org/10.1103/PhysRevLett.130.236401} {\bibfield  {journal} {\bibinfo  {journal} {Phys Rev Lett}\ }\textbf {\bibinfo {volume} {130}},\ \bibinfo {pages} {236401} (\bibinfo {year} {2023})}\BibitemShut {NoStop}%
\bibitem [{\citenamefont {Fishman}\ \emph {et~al.}(2022)\citenamefont {Fishman}, \citenamefont {White},\ and\ \citenamefont {Stoudenmire}}]{itensor}%
  \BibitemOpen
  \bibfield  {author} {\bibinfo {author} {\bibfnamefont {M.}~\bibnamefont {Fishman}}, \bibinfo {author} {\bibfnamefont {S.~R.}\ \bibnamefont {White}},\ and\ \bibinfo {author} {\bibfnamefont {E.~M.}\ \bibnamefont {Stoudenmire}},\ }\bibfield  {title} {\bibinfo {title} {{The ITensor Software Library for Tensor Network Calculations}},\ }\href {https://doi.org/10.21468/SciPostPhysCodeb.4} {\bibfield  {journal} {\bibinfo  {journal} {SciPost Phys Codebases}\ ,\ \bibinfo {pages} {4}} (\bibinfo {year} {2022})}\BibitemShut {NoStop}%
\bibitem [{\citenamefont {Moller}\ \emph {et~al.}(2008)\citenamefont {Moller}, \citenamefont {Madsen},\ and\ \citenamefont {Molmer}}]{RydbergBlockade}%
  \BibitemOpen
  \bibfield  {author} {\bibinfo {author} {\bibfnamefont {D.}~\bibnamefont {Moller}}, \bibinfo {author} {\bibfnamefont {L.}~\bibnamefont {Madsen}},\ and\ \bibinfo {author} {\bibfnamefont {K.}~\bibnamefont {Molmer}},\ }\bibfield  {title} {{\selectlanguage {English}\bibinfo {title} {Quantum gates and multiparticle entanglement by rydberg excitation blockade and adiabatic passage}},\ }\href {https://doi.org/10.1103/PhysRevLett.100.170504}{\bibfield  {journal} {\bibinfo  {journal} {Phys Rev Lett}\ }\textbf {\bibinfo {volume} {100}},\  \bibinfo {pages} {170504} (\bibinfo {year} {2008})}\BibitemShut {NoStop}%
\bibitem [{\citenamefont {Miles}\ \emph {et~al.}(2023)\citenamefont {Miles}, \citenamefont {Samajdar}, \citenamefont {Ebadi}, \citenamefont {Wang}, \citenamefont {Pichler}, \citenamefont {Sachdev}, \citenamefont {Lukin}, \citenamefont {Greiner}, \citenamefont {Weinberger},\ and\ \citenamefont {Kim}}]{PhaseTransition}%
  \BibitemOpen
  \bibfield  {author} {\bibinfo {author} {\bibfnamefont {C.}~\bibnamefont {Miles}}, \bibinfo {author} {\bibfnamefont {R.}~\bibnamefont {Samajdar}}, \bibinfo {author} {\bibfnamefont {S.}~\bibnamefont {Ebadi}}, \bibinfo {author} {\bibfnamefont {T.~T.}\ \bibnamefont {Wang}}, \bibinfo {author} {\bibfnamefont {H.}~\bibnamefont {Pichler}}, \bibinfo {author} {\bibfnamefont {S.}~\bibnamefont {Sachdev}}, \bibinfo {author} {\bibfnamefont {M.~D.}\ \bibnamefont {Lukin}}, \bibinfo {author} {\bibfnamefont {M.}~\bibnamefont {Greiner}}, \bibinfo {author} {\bibfnamefont {K.~Q.}\ \bibnamefont {Weinberger}},\ and\ \bibinfo {author} {\bibfnamefont {E.-A.}\ \bibnamefont {Kim}},\ }\bibfield  {title} {\bibinfo {title} {Machine learning discovery of new phases in programmable quantum simulator snapshots},\ }\href {https://doi.org/10.1103/PhysRevResearch.5.013026} {\bibfield  {journal} {\bibinfo  {journal} {Phys Rev Res}\ }\textbf {\bibinfo {volume} {5}},\ \bibinfo {pages} {013026} (\bibinfo {year} {2023})}\BibitemShut {NoStop}%
\bibitem [{\citenamefont {Merali}\ \emph {et~al.}(2024)\citenamefont {Merali}, \citenamefont {De~Vlugt},\ and\ \citenamefont {Melko}}]{merali_stochastic_2021}%
  \BibitemOpen
  \bibfield  {author} {\bibinfo {author} {\bibfnamefont {E.}~\bibnamefont {Merali}}, \bibinfo {author} {\bibfnamefont {I.~J.~S.}\ \bibnamefont {De~Vlugt}},\ and\ \bibinfo {author} {\bibfnamefont {R.~G.}\ \bibnamefont {Melko}},\ }\bibfield  {title} {\bibinfo {title} {Stochastic series expansion quantum monte carlo for rydberg arrays},\ }\href {https://doi.org/10.21468/SciPostPhysCore.7.2.016} {\bibfield  {journal} {\bibinfo  {journal} {SciPost Phys Core}\ }\textbf {\bibinfo {volume} {7}},\ \bibinfo {pages} {016} (\bibinfo {year} {2024})}\BibitemShut {NoStop}%
\bibitem[{\citenamefont{Zerweck}\ and\
\citenamefont{Carrasquilla}(2026)}]{Carrasquilla2026}%
  \BibitemOpen
  \bibfield{author}{%
    \bibinfo{author}{\bibfnamefont{T.}~\bibnamefont{Zerweck}}\
    and\ \bibinfo{author}{\bibfnamefont{J.}~\bibnamefont{Carrasquilla}},\ }%
  \bibfield{title}{\bibinfo{title}{%
    QMBL: the Quantum Many-Body Leaderboard.
    A live record of published ground-state energies for lattice
    Hamiltonians},\ }%
  \href{https://doi.org/10.48550/arXiv.2610.22387}{%
    \bibfield{journal}{%
      \bibinfo{journal}{arXiv:2610.22387 [cond-mat.str-el]}\ }%
    (\bibinfo{year}{2026})}%
  \BibitemShut{NoStop}%
\bibitem [{\citenamefont {Rende}\ \emph {et~al.}(2026)\citenamefont {Rende}, \citenamefont {Sinibaldi}, \citenamefont {Viteritti}, \citenamefont {Wiersema}, \citenamefont {Georges},\ and\ \citenamefont {Carleo}}]{rende2026scalinglawsneuralnetworkquantum}%
  \BibitemOpen
  \bibfield  {author} {\bibinfo {author} {\bibfnamefont {R.}~\bibnamefont {Rende}}, \bibinfo {author} {\bibfnamefont {A.}~\bibnamefont {Sinibaldi}}, \bibinfo {author} {\bibfnamefont {L.~L.}\ \bibnamefont {Viteritti}}, \bibinfo {author} {\bibfnamefont {R.}~\bibnamefont {Wiersema}}, \bibinfo {author} {\bibfnamefont {A.}~\bibnamefont {Georges}},\ and\ \bibinfo {author} {\bibfnamefont {G.}~\bibnamefont {Carleo}},\ }\bibinfo {title} {Scaling laws for neural-network quantum states,\ } \href {https://doi.org/10.48550/arXiv.2606.02794} {\bibfield{journal}{\bibinfo{journal}{arXiv:2606.02794[cond-mat.dis-nn]}} (\bibinfo {year} {2026})} \BibitemShut {NoStop}%
\bibitem [{\citenamefont {Sharma}\ \emph {et~al.}(2026)\citenamefont {Sharma}, \citenamefont {Shokry}, \citenamefont {Nutakki}, \citenamefont {Simard}, \citenamefont {Ferrero},\ and\ \citenamefont {Vicentini}}]{s3kd-bc4q}%
  \BibitemOpen
  \bibfield  {author} {\bibinfo {author} {\bibfnamefont {L.}~\bibnamefont {Sharma}}, \bibinfo {author} {\bibfnamefont {A.}~\bibnamefont {Shokry}}, \bibinfo {author} {\bibfnamefont {R.}~\bibnamefont {Nutakki}}, \bibinfo {author} {\bibfnamefont {O.}~\bibnamefont {Simard}}, \bibinfo {author} {\bibfnamefont {M.}~\bibnamefont {Ferrero}},\ and\ \bibinfo {author} {\bibfnamefont {F.}~\bibnamefont {Vicentini}},\ }\bibfield  {title} {\bibinfo {title} {Comparing symmetrized determinant neural quantum states for the hubbard model},\ }\href {https://doi.org/10.1103/s3kd-bc4q} {\bibfield  {journal} {\bibinfo  {journal} {Phys Rev B}\ }\textbf {\bibinfo {volume} {113}},\ \bibinfo {pages} {245104} (\bibinfo {year} {2026})}\BibitemShut {NoStop}%
\bibitem [{\citenamefont {Paszke}\ \emph {et~al.}(2019)\citenamefont {Paszke}, \citenamefont {Gross}, \citenamefont {Massa}, \citenamefont {Lerer}, \citenamefont {Bradbury}, \citenamefont {Chanan}, \citenamefont {Killeen}, \citenamefont {Lin}, \citenamefont {Gimelshein}, \citenamefont {Antiga}, \citenamefont {Desmaison}, \citenamefont {Kopf}, \citenamefont {Yang}, \citenamefont {DeVito}, \citenamefont {Raison}, \citenamefont {Tejani}, \citenamefont {Chilamkurthy}, \citenamefont {Steiner}, \citenamefont {Fang}, \citenamefont {Bai},\ and\ \citenamefont {Chintala}}]{cite_pytorch}%
  \BibitemOpen
  \bibfield  {author} {\bibinfo {author} {\bibfnamefont {A.}~\bibnamefont {Paszke}}, \bibinfo {author} {\bibfnamefont {S.}~\bibnamefont {Gross}}, \bibinfo {author} {\bibfnamefont {F.}~\bibnamefont {Massa}}, \bibinfo {author} {\bibfnamefont {A.}~\bibnamefont {Lerer}}, \bibinfo {author} {\bibfnamefont {J.}~\bibnamefont {Bradbury}}, \bibinfo {author} {\bibfnamefont {G.}~\bibnamefont {Chanan}}, \bibinfo {author} {\bibfnamefont {T.}~\bibnamefont {Killeen}}, \bibinfo {author} {\bibfnamefont {Z.}~\bibnamefont {Lin}}, \bibinfo {author} {\bibfnamefont {N.}~\bibnamefont {Gimelshein}}, \bibinfo {author} {\bibfnamefont {L.}~\bibnamefont {Antiga}}, \bibinfo {author} {\bibfnamefont {A.}~\bibnamefont {Desmaison}}, \bibinfo {author} {\bibfnamefont {A.}~\bibnamefont {Kopf}}, \bibinfo {author} {\bibfnamefont {E.}~\bibnamefont {Yang}}, \bibinfo {author} {\bibfnamefont {Z.}~\bibnamefont {DeVito}}, \bibinfo {author} {\bibfnamefont {M.}~\bibnamefont {Raison}}, \bibinfo {author} {\bibfnamefont {A.}~\bibnamefont {Tejani}}, \bibinfo
  {author} {\bibfnamefont {S.}~\bibnamefont {Chilamkurthy}}, \bibinfo {author} {\bibfnamefont {B.}~\bibnamefont {Steiner}}, \bibinfo {author} {\bibfnamefont {L.}~\bibnamefont {Fang}}, \bibinfo {author} {\bibfnamefont {J.}~\bibnamefont {Bai}},\ and\ \bibinfo {author} {\bibfnamefont {S.}~\bibnamefont {Chintala}},\ }\bibfield  {title} {\bibinfo {title} {Pytorch: An imperative style, high-performance deep learning library},\ }in\ \href {https://proceedings.neurips.cc/paper_files/paper/2019/file/bdbca288fee7f92f2bfa9f7012727740-Paper.pdf} {\emph {\bibinfo {booktitle} {Adv Neural Inf Process Syst}}},\ Vol.~\bibinfo {volume} {32},\ \bibinfo {editor} {edited by\ \bibinfo {editor} {\bibfnamefont {H.}~\bibnamefont {Wallach}}, \bibinfo {editor} {\bibfnamefont {H.}~\bibnamefont {Larochelle}}, \bibinfo {editor} {\bibfnamefont {A.}~\bibnamefont {Beygelzimer}}, \bibinfo {editor} {\bibfnamefont {F.}~\bibnamefont {d\textquotesingle Alch\'{e}-Buc}}, \bibinfo {editor} {\bibfnamefont {E.}~\bibnamefont
  {Fox}},\ and\ \bibinfo {editor} {\bibfnamefont {R.}~\bibnamefont {Garnett}}}\ (\bibinfo  {publisher} {Curran Associates, Inc.},\ \bibinfo {year} {2019})\BibitemShut {NoStop}%
\bibitem [{\citenamefont {Kidger}(2022)}]{cite_jaxtyping}%
  \BibitemOpen
  \bibfield  {author} {\bibinfo {author} {\bibfnamefont {P.}~\bibnamefont {Kidger}},\ }\href {https://github.com/patrick-kidger/jaxtyping} {\bibinfo {title} {jaxtyping: Type annotations and runtime checking for shape and dtype of jax/numpy/pytorch/etc. arrays}},\ (\bibinfo {year} {2022})\BibitemShut {NoStop}%
\bibitem [{\citenamefont {da~Costa-Luis}\ \emph {et~al.}()\citenamefont {da~Costa-Luis}, \citenamefont {Larroque}, \citenamefont {Altendorf}, \citenamefont {Mary}, \citenamefont {Korobov}, \citenamefont {Yorav-Raphael}, \citenamefont {Ivanov}, \citenamefont {Bargull}, \citenamefont {Rodrigues}, \citenamefont {CHEN}, \citenamefont {Newey}, \citenamefont {James}, \citenamefont {Zugnoni},\ and\ \citenamefont {Pagel}}]{cite_tqdm}%
  \BibitemOpen
  \bibfield  {author} {\bibinfo {author} {\bibfnamefont {C.}~\bibnamefont {da~Costa-Luis}}, \bibinfo {author} {\bibfnamefont {S.~K.}\ \bibnamefont {Larroque}}, \bibinfo {author} {\bibfnamefont {K.}~\bibnamefont {Altendorf}}, \bibinfo {author} {\bibfnamefont {H.}~\bibnamefont {Mary}}, \bibinfo {author} {\bibfnamefont {M.}~\bibnamefont {Korobov}}, \bibinfo {author} {\bibfnamefont {N.}~\bibnamefont {Yorav-Raphael}}, \bibinfo {author} {\bibfnamefont {I.}~\bibnamefont {Ivanov}}, \bibinfo {author} {\bibfnamefont {M.}~\bibnamefont {Bargull}}, \bibinfo {author} {\bibfnamefont {N.}~\bibnamefont {Rodrigues}}, \bibinfo {author} {\bibfnamefont {G.}~\bibnamefont {CHEN}}, \bibinfo {author} {\bibfnamefont {C.}~\bibnamefont {Newey}}, \bibinfo {author} {\bibnamefont {James}}, \bibinfo {author} {\bibfnamefont {M.}~\bibnamefont {Zugnoni}},\ and\ \bibinfo {author} {\bibnamefont {Pagel}},\ }\href {https://doi.org/10.5281/zenodo.22705034} {\bibinfo {title} {TQDM: A fast, extensible progress bar for python and CLI}},\ \BibitemShut {NoStop}%
\bibitem [{\citenamefont {Harris}\ \emph {et~al.}(2020)\citenamefont {Harris}, \citenamefont {Millman}, \citenamefont {van~der Walt}, \citenamefont {Gommers}, \citenamefont {Virtanen}, \citenamefont {Cournapeau}, \citenamefont {Wieser}, \citenamefont {Taylor}, \citenamefont {Berg}, \citenamefont {Smith}, \citenamefont {Kern}, \citenamefont {Picus}, \citenamefont {Hoyer}, \citenamefont {van Kerkwijk}, \citenamefont {Brett}, \citenamefont {Haldane}, \citenamefont {del R{\'\i}o}, \citenamefont {Wiebe}, \citenamefont {Peterson}, \citenamefont {G{\'e}rard-Marchant}, \citenamefont {Sheppard}, \citenamefont {Reddy}, \citenamefont {Weckesser}, \citenamefont {Abbasi}, \citenamefont {Gohlke},\ and\ \citenamefont {Oliphant}}]{cite_numpy}%
  \BibitemOpen
  \bibfield  {author} {\bibinfo {author} {\bibfnamefont {C.~R.}\ \bibnamefont {Harris}}, \bibinfo {author} {\bibfnamefont {K.~J.}\ \bibnamefont {Millman}}, \bibinfo {author} {\bibfnamefont {S.~J.}\ \bibnamefont {van~der Walt}}, \bibinfo {author} {\bibfnamefont {R.}~\bibnamefont {Gommers}}, \bibinfo {author} {\bibfnamefont {P.}~\bibnamefont {Virtanen}}, \bibinfo {author} {\bibfnamefont {D.}~\bibnamefont {Cournapeau}}, \bibinfo {author} {\bibfnamefont {E.}~\bibnamefont {Wieser}}, \bibinfo {author} {\bibfnamefont {J.}~\bibnamefont {Taylor}}, \bibinfo {author} {\bibfnamefont {S.}~\bibnamefont {Berg}}, \bibinfo {author} {\bibfnamefont {N.~J.}\ \bibnamefont {Smith}}, \bibinfo {author} {\bibfnamefont {R.}~\bibnamefont {Kern}}, \bibinfo {author} {\bibfnamefont {M.}~\bibnamefont {Picus}}, \bibinfo {author} {\bibfnamefont {S.}~\bibnamefont {Hoyer}}, \bibinfo {author} {\bibfnamefont {M.~H.}\ \bibnamefont {van Kerkwijk}}, \bibinfo {author} {\bibfnamefont {M.}~\bibnamefont {Brett}}, \bibinfo {author} {\bibfnamefont
  {A.}~\bibnamefont {Haldane}}, \bibinfo {author} {\bibfnamefont {J.~F.}\ \bibnamefont {del R{\'\i}o}}, \bibinfo {author} {\bibfnamefont {M.}~\bibnamefont {Wiebe}}, \bibinfo {author} {\bibfnamefont {P.}~\bibnamefont {Peterson}}, \bibinfo {author} {\bibfnamefont {P.}~\bibnamefont {G{\'e}rard-Marchant}}, \bibinfo {author} {\bibfnamefont {K.}~\bibnamefont {Sheppard}}, \bibinfo {author} {\bibfnamefont {T.}~\bibnamefont {Reddy}}, \bibinfo {author} {\bibfnamefont {W.}~\bibnamefont {Weckesser}}, \bibinfo {author} {\bibfnamefont {H.}~\bibnamefont {Abbasi}}, \bibinfo {author} {\bibfnamefont {C.}~\bibnamefont {Gohlke}},\ and\ \bibinfo {author} {\bibfnamefont {T.~E.}\ \bibnamefont {Oliphant}},\ }\bibfield  {title} {\bibinfo {title} {Array programming with {NumPy}},\ }\href {https://doi.org/10.1038/s41586-020-2649-2} {\bibfield  {journal} {\bibinfo  {journal} {Nature}\ }\textbf {\bibinfo {volume} {585}},\ \bibinfo {pages} {357} (\bibinfo {year} {2020})}\BibitemShut {NoStop}%
\bibitem [{\citenamefont {Hunter}(2007)}]{cite_matplotlib}%
  \BibitemOpen
  \bibfield  {author} {\bibinfo {author} {\bibfnamefont {J.~D.}\ \bibnamefont {Hunter}},\ }\bibfield  {title} {\bibinfo {title} {Matplotlib: A 2d graphics environment},\ }\href {https://doi.org/10.1109/MCSE.2007.55} {\bibfield  {journal} {\bibinfo  {journal} {Comp Sci Eng}\ }\textbf {\bibinfo {volume} {9}},\ \bibinfo {pages} {90} (\bibinfo {year} {2007})}\BibitemShut {NoStop}%
\bibitem [{\citenamefont {Vink}\ \emph {et~al.}(2025)\citenamefont {Vink} \emph {et~al.}}]{cite_polars}%
  \BibitemOpen
  \bibfield  {author} {\bibinfo {author} {\bibfnamefont {R.}\ \bibnamefont {Vink}}, \bibinfo {author} {\bibfnamefont {A.}\ \bibnamefont {Beedie}}, \bibinfo {author} {\bibfnamefont {O.}\ \bibnamefont {Peters}}, \bibinfo {author} {\bibfnamefont {G.}~\bibnamefont {Burghoorn}}, \bibinfo {author} {\bibfnamefont {S.}~\bibnamefont {de Gooijer}}, \bibinfo {author} {\bibfnamefont {M.~E.}~\bibnamefont {Gorelli}}, \bibinfo {author} {\bibfnamefont {A.}~\bibnamefont {Sprenkels}}, \bibinfo {author} {\bibfnamefont {K.}~\bibnamefont {Denecker}}, \bibinfo {author} {\bibfnamefont {J.}~\bibnamefont {van Zundert}}, \bibinfo {author} {\bibfnamefont {G.}\ \bibnamefont {Hulselmans}}, \bibinfo {author} {\bibfnamefont {C.}~\bibnamefont {Grinstead}}, \bibinfo {author} {\bibfnamefont {K.}~\bibnamefont {Patyk}}, \bibinfo {author} {\bibfnamefont {L.}~\bibnamefont {Manley}}, \bibinfo {author} {\bibfnamefont {K.}\ \bibnamefont {Valtar}}, \bibinfo {author} {\bibfnamefont {D.}~\bibnamefont {Szilagyi}}, \bibinfo {author} {\bibfnamefont
  {I.}~\bibnamefont {Turner-Trauring}}, \bibinfo {author} {\bibfnamefont {L.}\ \bibnamefont {Mitchell}}, \bibinfo {author} {\bibfnamefont {A.}~\bibnamefont {Afroozeh}}, \bibinfo {author} {\bibfnamefont {O.}~\bibnamefont {Borchert}}, \bibinfo {author} {\bibfnamefont {K.}~\bibnamefont {Genockey}}, \bibinfo {author} {\bibfnamefont {H.}~\bibnamefont {Harbeck}},\ and\ \bibinfo {author} {\bibfnamefont {L.}~\bibnamefont {Bergdoll}},\ }\href {https://zenodo.org/records/23187389} {\bibinfo {title} {Polars: Dataframes for the new era}} (\bibinfo {year} {2025})\BibitemShut {NoStop}%
\bibitem [{\citenamefont {{Faith O. Oyedemi}}(2026)}]{code_repo}%
  \BibitemOpen
  \bibfield  {author} {\bibinfo {author} {\bibnamefont {{Faith O. Oyedemi}}},\ }\href {https://github.com/APRIQuOt/mlp-mixer-wave-function} {\bibfield {title} {\bibinfo {title} {MLP-mixer-wave-function}}},\ \bibinfo {howpublished} {\url{https://github.com/APRIQuOt/mlp-mixer-wave-function}} (\bibinfo {year} {2026})\BibitemShut {NoStop}%
\bibitem [{\citenamefont {{Hendrycks}}\ and\ \citenamefont {{Gimpel}}(2016)}]{GELU}%
  \BibitemOpen
  \bibfield  {author} {\bibinfo {author} {\bibfnamefont {D.}~\bibnamefont {{Hendrycks}}}\ and\ \bibinfo {author} {\bibfnamefont {K.}~\bibnamefont {{Gimpel}}},\ }\bibfield  {title} {\bibinfo {title} {{Gaussian Error Linear Units (GELUs)}},\ }\href {https://doi.org/10.48550/arXiv.1606.08415} {\bibfield  {journal} {\bibinfo  {journal} {arXiv:1606.08415[cs.LG]}} (\bibinfo {year} {2016})}\BibitemShut {NoStop}%
\bibitem [{\citenamefont {Liu}\ \emph {et~al.}(2022)\citenamefont {Liu}, \citenamefont {Li}, \citenamefont {Tao}, \citenamefont {Liang},\ and\ \citenamefont {Zheng}}]{LIUParadigmShift}%
  \BibitemOpen
  \bibfield  {author} {\bibinfo {author} {\bibfnamefont {R.}~\bibnamefont {Liu}}, \bibinfo {author} {\bibfnamefont {Y.}~\bibnamefont {Li}}, \bibinfo {author} {\bibfnamefont {L.}~\bibnamefont {Tao}}, \bibinfo {author} {\bibfnamefont {D.}~\bibnamefont {Liang}},\ and\ \bibinfo {author} {\bibfnamefont {H.-T.}\ \bibnamefont {Zheng}},\ }\bibfield  {title} {\bibinfo {title} {Are we ready for a new paradigm shift? a survey on visual deep mlp},\ }\href {https://doi.org/https://doi.org/10.1016/j.patter.2022.100520} {\bibfield  {journal} {\bibinfo  {journal} {Patterns}\ }\textbf {\bibinfo {volume} {3}},\ \bibinfo {pages} {100520} (\bibinfo {year} {2022})}\BibitemShut {NoStop}%
\bibitem [{\citenamefont {Zhang}\ and\ \citenamefont {Di~Ventra}(2023)}]{MultipurposeTransformer}%
  \BibitemOpen
  \bibfield  {author} {\bibinfo {author} {\bibfnamefont {Y.-H.}\ \bibnamefont {Zhang}}\ and\ \bibinfo {author} {\bibfnamefont {M.}~\bibnamefont {Di~Ventra}},\ }\bibfield  {title} {\bibinfo {title} {Transformer quantum state: A multipurpose model for quantum many-body problems},\ }\href {https://doi.org/10.1103/PhysRevB.107.075147} {\bibfield  {journal} {\bibinfo  {journal} {Phys Rev B}\ }\textbf {\bibinfo {volume} {107}},\ \bibinfo {pages} {075147} (\bibinfo {year} {2023})}\BibitemShut {NoStop}%
\bibitem [{\citenamefont {Germain}\ \emph {et~al.}(2015{\natexlab{b}})\citenamefont {Germain}, \citenamefont {Gregor}, \citenamefont {Murray},\ and\ \citenamefont {Larochelle}}]{MADE}%
  \BibitemOpen
  \bibfield  {author} {\bibinfo {author} {\bibfnamefont {M.}~\bibnamefont {Germain}}, \bibinfo {author} {\bibfnamefont {K.}~\bibnamefont {Gregor}}, \bibinfo {author} {\bibfnamefont {I.}~\bibnamefont {Murray}},\ and\ \bibinfo {author} {\bibfnamefont {H.}~\bibnamefont {Larochelle}},\ }\bibfield  {title} {\bibinfo {title} {MADE: Masked autoencoder for distribution estimation},\ }in\ \href {https://proceedings.mlr.press/v37/germain15.html} {\emph {\bibinfo {booktitle} {Proceedings of ICML 32}}},\  Vol.~\bibinfo {volume} {37},\ \bibinfo {editor} {edited by\ \bibinfo {editor} {\bibfnamefont {F.}~\bibnamefont {Bach}}\ and\ \bibinfo {editor} {\bibfnamefont {D.}~\bibnamefont {Blei}}}\ (\bibinfo  {publisher} {PMLR},\ \bibinfo {address} {Lille, France},\ \bibinfo {year} {2015})\ pp.\ \bibinfo {pages} {881--889}\BibitemShut {NoStop}%
\end{thebibliography}
\end{document}